% ============================================================================
% "K_L -> pi^0 e+ e- and B -> X_s e+ e- decay in the MSSM"
% ============================================================================
 
% ==========================================================================
% This file uses the Harvmac macros and should be printed out in
% "big" format.
% ==========================================================================

\input harvmac

% The following command eliminates black boxes that appear when
% equations run over RHS margins:
\overfullrule=0pt
 
% Small capital subscripts
\def\A{{\scriptscriptstyle A}}
\def\B{{\scriptscriptstyle B}}

\def\D{{\scriptscriptstyle D}}
\def\E{{\scriptscriptstyle E}}

\def\G{{\scriptscriptstyle G}}

\def\I{{\scriptscriptstyle I}}
\def\J{{\scriptscriptstyle J}}

\def\L{{\scriptscriptstyle L}}
\def\M{{\scriptscriptstyle M}}
\def\N{{\scriptscriptstyle N}}

\def\R{{\scriptscriptstyle R}}

\def\T{{\scriptscriptstyle T}}
\def\U{{\scriptscriptstyle U}}

\def\W{{\scriptscriptstyle W}}

\def\Y{{\scriptscriptstyle Y}}
\def\Z{{\scriptscriptstyle Z}}
 
% Calligraphy letters
 
\def\CA{{\cal A}}

\def\CH{{\cal H}}

\def\CL{{\cal L}}

% Greek letters

\def\b{\beta}
\def\d{\delta}

\def\th{\theta}
\def\u{\mu}
\def\v{\nu}

% Fractions

\def\half{{1 \over 2}}
\def\quarter{{1 \over 4}}

\def\third{{1 \over 3}}

\def\twothirds{{2 \over 3}}

\font\sy=cmmi10 scaled \magstep2

% Aliases

\def\aEM{\alpha_{\scriptscriptstyle EM}}
\def\aS{\alpha_s}
\def\at{\alpha_{\tilde t}}

\def\Athree{{\scriptscriptstyle A3}}
\def\bar#1{\overline{#1}}
\def\bigmu{\hbox{\sy\char'26}}

\def\Ai{{\scriptscriptstyle Ai}}
\def\Bi{{\scriptscriptstyle Bi}}
\def\Br{{\rm Br}}
\def\Bsg{B \to X_s \gamma}

\def\Bsee{B \to X_s e^+ e^-}

\def\Bsmm{B \to X_s \mu^+ \mu^-}
\def\bsll{b \to s \ell^+ \ell^-}
\def\Bsll{B \to X_s \ell^+ \ell^-}
\def\ccdot{\hbox{\kern-.1em$\cdot$\kern-.1em}}
 % Slashed covariant deriv
		% \not{CP}
\def\dash{{\> \over \>}} 		% Hyphen in equations
\def\dAB{\delta_{\A\B}}
\def\dbar{\bar{d}}
\def\dIJ{\delta_{\I\J}}
\def\DL{{D_\L}}
\def\DR{{D_\R}}
\def\Dslash{D\hskip-0.65 em / \hskip+0.30 em} % Slashed covariant deriv
\def\ebar{\bar{e}}
\def\eff{{\rm eff}}
\def\EL{{E_\L}}
\def\ER{{E_\R}}

\def\GeV{\>\, \rm GeV}
\def\GF{G_{\scriptscriptstyle F}}
\def\gfive{\gamma^5}
\def\gtap{\raise.3ex\hbox{$>$\kern-.75em\lower1ex\hbox{$\sim$}}}
\def\hc{{\rm h.c.}}
\def\Heff{\CH_{\rm eff}}
\def\Im{{\rm Im}}

\def\Jpsi{{J/\psi}}

\def\KMd{K^*_{ts} K_{td}}
\def\KMangles{K^*_{ts} K_{tq}}
\def\KManglesb{K^*_{ts} K_{tb}}
\def\kpiee{K_\L \to \pi^0 e^+ e^-}
\def\L{{\scriptscriptstyle L}}

\def\Li{{\rm Li}}
\def\ltap{\raise.3ex\hbox{$<$\kern-.75em\lower1ex\hbox{$\sim$}}}

\def\mb{{m_b}}
\def\mc{{m_c}}
\def\mchI{m_{\tilde{\chi}^\pm_\I}}
\def\mchJ{m_{\tilde{\chi}^\pm_\J}}
\def\mdsqA{m_{\tilde{d}_A}}
\def\mdsqB{m_{\tilde{d}_B}}
\def\meslone{m_{\tilde{e}_1}}
\def\meslfour{m_{\tilde{e}_4}}

\def\mgluino{m_{\tilde{g}}}
\def\MGUT{{M_{\rm GUT}}}
\def\mhsq{m^2_{h^\pm}}
\def\mneI{m_{\tilde{\chi}^0_\I}}
\def\mneJ{m_{\tilde{\chi}^0_\J}}
\def\MSSM{{\rm MSSM}}
\def\msneutrinoone{m_{\tilde{\nu}_1}}
\def\mt{{m_t}}
\def\mtsq{m_t^2}
\def\musqA{m_{\tilde{u}_A}}
\def\musqB{m_{\tilde{u}_B}}
\def\mW{m_\W}
\def\mWsq{m_\W^2}
\def\mZ{m_\Z}
\def\mZsq{m_\Z^2}
\def\NL{{N_\L}}

\def\R{{\scriptscriptstyle R}}
\def\Pminus{P_-}
\def\Pplus{P_+}

\def\Re{{\rm Re}}
\def\rhokpiee{\rho(\kpiee)}
\def\sbar{\bar{s}}

\def\shat{{\hat s}}
\def\sinsq{\sin^2 \theta }
  % Macro for slashes thru vectors.
\def\SM{{\rm SM}}
\def\sp{\>\>}
\def\SUSY{{\rm SUSY}}
\def\therefore{{\hbox{..}\kern-.43em \raise.5ex \hbox{.}}\>\>}
\def\tildeW{\widetilde W}
\def\twoA{{\scriptscriptstyle 2A}}
\def\xt{x_t}

\def\UL{{U_\L}}
\def\UR{{U_\R}}

% Style-sensitive Poor-Man's-Bold command, produces bold greek letters.
% Usage $ ... \pmb\gamma ... $
% Adapted from TeXbook p386 (\pmb) and p360 (\mathpallette)
\newdimen\pmboffset
\pmboffset 0.022em
\def\oldpmb#1{\setbox0=\hbox{#1}%
 \copy0\kern-\wd0
 \kern\pmboffset\raise 1.732\pmboffset\copy0\kern-\wd0
 \kern\pmboffset\box0}
\def\pmb#1{\mathchoice{\oldpmb{$\displaystyle#1$}}{\oldpmb{$\textstyle#1$}}
      {\oldpmb{$\scriptstyle#1$}}{\oldpmb{$\scriptscriptstyle#1$}}}

% Modified Appendix definition with no unwanted letter after the boldface
% word APPENDIX.
%
%\def\appendix#1#2{\global\meqno=1\global\subsecno=0\xdef\secsym{\hbox{#1.}}
%\bigbreak\bigskip\noindent{\bf Appendix. #2}\message{(#1. #2)}
%\writetoca{Appendix {#1.} {#2}}\par\nobreak\medskip\nobreak}
 
% ==========================================================================
% References
% ==========================================================================

\nref\EWcorr{J. Ellis, G. L. Fogli and E. Lisi, 
             Phys. Lett. {\bf B324} (1994) 173; {\bf B333} (1994) 118\semi
   P.H. Chankowski and S. Pokorski, Phys. Lett. {\bf B366} (1996) 188\semi
   P. Langacker, NSF-ITP-95-140 (1995) (hep-ph/9511207).} 
\nref\SUSYFCNC{ 
B. de Carlos and J.A. Casas, Phys. Lett. {\bf B349} (1995) 300\semi
V. Barger, M.S. Berger, P. Ochmann and R.J.N. Phillips,
                               Phys. Rev {\bf D51} (1995) 2438\semi
E. Gabrielli, A Masiero and L. Silvestrini, ROM2F/95/23 (1995) 
 (hep-ph/9510215)\semi
G.C. Cho, Y. Kizukuri and N. Oshimo, TKU-HEP 95/02 (1995) (hep-ph/9509277).}
\nref\CLEO{M.S. Amal {\it et al.} (CLEO Collaboration), Phys. Rev. Lett.
  {\bf 74} (1995) 2885.}
\nref\BG{R. Barbieri and G.F. Giudice, Phys. Lett. {\bf B309} (1993) 86\semi
         R. Garisto and J.N. Ng, Phys. Lett. {\bf 315} (1993) 372.}
\nref\PDB{Review of particle properties, Phys. Rev. {\bf D50} (1994) 1173.}
\nref\MI{B. Winstein and L. Wolfenstein, Rev. Mod. Phys. 
  {\bf 65} (1993) 1113\semi
K. Arisaka et al., ``KAMI conceptual design report'', Fermilab FN-568 
 (1991)\semi
Y. Wah, private communication.}
\nref\CLEOII{R. Balest {\it et al.} (CLEO Collaboration), CLEO-CONF 94-4 
 (1994).}
\nref\CDF{C. Anway-Wiese {\it et al.} (CDF Collaboration), 
 Fermilab-Conf-95/201-E (1995).}
\nref\Flynn{J.M. Flynn and L. Randall, Nucl. Phys. {\bf B326} (1989) 31.}
\nref\Dib{C.O. Dib, I. Dunietz and F.J. Gilman, Phys. Rev. {\bf D39} (1989) 
  2639.}
\nref\Buras{A.J. Buras, M.E. Lautenbacher, M. Misiak and M. M\"unz, Nucl.
 Phys. {\bf B423} (1994) 349.}
\nref\Misiak{M.~Misiak, Nucl. Phys. {\bf B393} (1993) 23, 
	     (E) {\bf B439} (1995) 461\semi
	     A.J. Buras and M. M\"unz, Phys. Rev. {\bf D52} (1995) 186.}
\nref\Cohen{A.G. Cohen, G. Ecker and A. Pich, Phys. Lett. {\bf B304} (1993) 
 347.}
\nref\Ecker{G. Ecker, A. Pich and E. de Rafael, Nucl. Phys. {\bf B303} (1988) 
 665\semi
 A. Pich, Introduction to Chiral Perturbation Theory, {\it in} Proc. of the 
 Fifth Mexican School of Particles and Fields 1992, ed. J.L. Lucio M. and M. 
 Vargas (American Institute of Physics, 1994) p. 95.}
\nref\Donoghue{J.F. Donoghue and F. Gabbiani, Phys. Rev. {\bf D51} (1995) 
 2187.}
\nref\Bertolini{S. Bertolini, F. Borzumati, A. Masiero and G. Ridolfi, Nucl. 
 Phys. {\bf B353} (1991) 591.}
\nref\Ali{A. Ali, G.F. Giudice and T. Mannel, Z. Phys. {\bf C67} (1995) 417.}
\nref\HaberKane{H.E. Haber and G.L. Kane, Phys. Rep. {\bf 117} (1985) 75.}
\nref\Haber{H.E. Haber, Introductory Low-Energy Supersymmetry, {\it in} Proc. 
 of the Theoretical Advanced Study Institute 1992, ed. J. Harvey and J. 
 Polchinsky (World Scientific, Singapore, 1993) p.~827.}
\nref\Derendinger{J.-P. Derendinger, Lecture Notes on Globally Supersymmetric 
 Theories in Four and Two Dimensions, {\it in} Proc. of the Hellenic School of 
 Elementary Particle Physics 1989, ed. E.N. Argyres, N. Tracas and G. Zoupanos
 (World Scientific, Singapore, 1990) p.~790.}
\nref\Ibanez{L. Ibanez and G. Ross, Phys. Lett. {\bf B110} (1982) 215\semi
  L. Ibanez, Nucl. Phys. {\bf B218} (1983) 514.}
\nref\Inoue{K. Inoue {\it et al}, Prog. Theory. Phys. {\bf 68} (1982) 927.}
\nref\Wise{L. Alvarez-Guame, M. Claudson and M. Wise, Nucl. Phys. {\bf 
  B207} (1982) 96.}
\nref\Rosiek{J. Rosiek, Phys. Rev. {\bf D41} (1990) 3464.}
\nref\SUGRA{A.H. Chamseddine, R. Arnowitt and P. Nath, 
                                Phys. Rev. Lett. {\bf 49} (1982) 970\semi
            S. Soni and A. Weldon, Phys. Lett. {\bf B126} (1983) 215\semi
L. Hall, J. Lykken and S. Weinberg, Phys. Rev. {\bf D27} (1983) 2359\semi
R. Barbieri, S. Ferrara and C.A. Savoy, Phys. Lett. {\bf B119} (1983) 343\semi
H.P. Nilles, M. Srednicki and D. Wyler, Phys. Lett. {\bf B120} (1983) 346\semi
R. Barbieri, J.Louis and M. Moreti, Phys. Lett. {\bf B312} (1993) 451.}
\nref\Buchmuller{W. Buchm\"uller and D. Wyler,  Phys. Lett. {\bf B121} (1983) 
 321\semi
J. Polchinski and M.B. Wise, Phys. Lett. {\bf B125} (1983) 393.}
\nref\LEPx{L. Rolandi (ALEPH), H. Dijkstra (OPAL), D. Strickland (L3)
and G. Wilson (OPAL), Joint Seminar on the First Results from LEP 1.5,
CERN, December 12, 1995.}
\nref\BMMP{A.~J.~Buras, M.~Misiak, M.~M\"unz and S.~Pokorski,
 Nucl. Phys. {\bf B424} (1994) 374.}
\nref\GSW{B. Grinstein, M.J. Savage and M.B. Wise, Nucl. Phys. {\bf B319}, 
  (1989) 271.}
\nref\Falk{A.F. Falk, M. Luke and M.J. Savage, Phys. Rev. {\bf D49} (1994) 
 3367.}
\nref\Ferrara{S. Ferrara and E. Remiddi, Phys. Lett. {\bf B53} (1974) 347.}
\nref\Ligeti{Z. Ligeti and M. Wise, CALT-68-2029 (1995), hep-ph 9512225.}

% ==========================================================================
% Figure captions
% ==========================================================================
 
\nfig\MSSMgraphs{One-loop MSSM (a) bubble, (b) penguin and (c) box graphs 
which contribute to the short distance coefficients of effective theory 
operators that mediate $\kpiee$ and $\Bsll$ decay.} 
\nfig\kpieeLEGOplots{Fractional suppression $\rhokpiee$ of the direct CP 
violating component of the $\kpiee$ branching fraction in the MSSM relative to 
the Standard Model prediction.  The suppression factor is plotted in (a) as a 
function of $A_0(\MGUT)$ and $m_0(\MGUT)$ with $m_{\tildeW}(\mZ) = 90 
\GeV$ and $\tan\beta=5$ held fixed.  The same quantity is displayed in (b) as a 
function of $m_{\tildeW}(\mZ)$ and $\tan\beta$ with $A_0(\MGUT)= -500 
\GeV$ and $m_0(\MGUT)=100 \GeV$.  Excluded MSSM parameter space points are 
indicated in both plots by LEGO blocks saturated at their maximum values.}
\nfig\AsymLEGOplot{Ratio of the nonresonant electron asymmetry observable 
$\CA(\Bsee)_{\rm NR}$ calculated in the MSSM relative to its Standard Model 
value.  The ratio is plotted as a function of $A_0(\MGUT)$ and $m_0(\MGUT)$ 
with $m_{\tildeW}(\mZ) = 150 \GeV$ and $\tan\beta=20$ held fixed.  
Excluded MSSM parameter space points are indicated in the plot by LEGO blocks 
saturated at their maximum values.}

% ==========================================================================
% Title page
% ==========================================================================

\def\CITTitle#1#2#3{\nopagenumbers\abstractfont
\hsize=\hstitle\rightline{#1}
\vskip 0.3in\centerline{\titlefont #2} \centerline{\titlefont #3}
\abstractfont\vskip .3in\pageno=0}
 
\CITTitle{{\baselineskip=10pt plus 1pt minus 1pt
  \vbox{\hbox{ZU-TH 24/95}
	\hbox{CALT-68-2028}
        \hbox{DOE RESEARCH AND}\hbox{DEVELOPMENT REPORT}
%         \hbox{hep-ph XXXXXX}
        }}}
{$K_L \to \pi^0 e^+ e^-$ and $B \to X_s \ell^+ \ell^-$ Decay}
{in the MSSM}
\centerline{
  Peter Cho\footnote{$^\dagger$}{Work supported in part by a DuBridge 
  Fellowship and by the U.S. Dept. of Energy under DOE Grant no. 
  DE-FG03-92-ER40701.}}
%\centerline{Lauritsen Laboratory}
\centerline{California Institute of Technology}
\centerline{Pasadena, CA  91125}

\smallskip
\centerline{and}
\smallskip

\centerline{Miko\l aj Misiak\footnote
{$^\ddagger$}{Work supported by Schweizerischer Nationalfonds.}\footnote
{$^*$}{Partially supported by the Committee for Scientific 
Research, Poland.}
and Daniel Wyler$^{\ddagger}$}

\centerline{Institut f\"ur Theoretische Physik der Universit\"at Z\"urich,}
\centerline{Winterthurerstrasse 190, CH-8057 Z\"urich}

\vskip 0.2in
\centerline{\bf Abstract}
\bigskip

	The flavor changing neutral current processes $K_L \to \pi^0 e^+ e^-$,
$B \to X_s e^+ e^-$ and $B \to X_s \mu^+ \mu^-$ are studied within the minimal
supersymmetric extension of the Standard Model.  We first examine the rates 
for these decay modes in the MSSM with a universal soft supersymmetry breaking 
sector at a Grand Unification scale.  We later relax the universality condition 
and investigate the FCNC transitions in a more general class of models with 
negligible flavor violation in squark mixing matrices.  We find that the MSSM 
prediction for the kaon channel's branching fraction differs from its Standard 
Model value by at most 30\% over the entire allowed parameter space.  On the 
other hand, supersymmetric contributions could potentially enhance certain
$B \to X_s \ell^+ \ell^-$ observables by more than 100\% relative to Standard 
Model expectations.  The impact of supersymmetry upon the $B$ meson modes is 
strongly correlated with the MSSM value for the Wilson coefficient of the 
magnetic moment operator that mediates $B \to X_s \gamma$.  

\Date{1/96}

% ==========================================================================
\newsec{Introduction}
% ==============================================================================

	The Minimal Supersymmetric Standard Model (MSSM) remains one of the 
few well-motivated extensions of the Standard Model which has survived 
precision electroweak measurements \EWcorr.  The need to subject this theory 
to other complementary experimental tests has consequently grown with time.  
Flavor Changing Neutral Current (FCNC) phenomenology represents one area where 
data are stringently confronting the MSSM \SUSYFCNC.  Deviations from the 
Standard Model may be observed in FCNC processes long before superpartners are
detected at high energy colliders.  Alternatively, failure to detect such 
departures places constraints upon weak scale supersymmetry and all other 
theories of physics beyond the Standard Model.  For example, the recent CLEO 
observation of inclusive $\Bsg$ decay rules out charged Higgs bosons lighter 
than 260 GeV in Two-Higgs Doublet models \CLEO.  This lower bound lies far 
beyond the reach of present direct searches. The $\Bsg$ measurement similarly 
constrains the MSSM, but its restrictive power is diminished by possible 
cancellations between different superparticle contributions \BG.  It is 
therefore important to study the sensitivity of other FCNC processes to 
supersymmetry and determine their limiting capabilities.

	In this article, we investigate the rare decays $\kpiee$, 
$\Bsee$ and $\Bsmm$ within the MSSM framework.  
Positive signals in these channels are expected to be observed within the next 
few years provided their rates do not lie significantly below Standard Model
predictions.  The current experimental upper bound on the first
process $\Br(\kpiee)_{\rm exp} < 4.3 \times 10^{-9}$
\PDB\ is three orders of magnitude larger than the anticipated
Standard Model branching fraction.  Yet this mode is expected to be
detected at Fermilab following completion of the Main Injector \MI.
Present CLEO and CDF exclusive limits on the second and third channels,
$\Br(B^0 \to K^{*0} e^+ e^-)_{\rm CLEO} < 1.6 \times
10^{-5}$ \CLEOII\ and $\Br(B^0 \to K^{*0} \mu^+ \mu^-)_{\rm CDF} < 2.1
\times 10^{-5}$ \CDF, lie within an order of magnitude of Standard
Model predictions.  Evidence for short distance $b \to s \ell^+
\ell^-$ decay may therefore soon be seen with the upgraded CLEO
detector and in the full Run Ib Tevatron data set.

	$\kpiee$ and $\Bsll$ decay have both been extensively studied in the 
Standard Model.  The impact of conventional QCD upon their rates can 
{\it a priori} be comparable to that from any new physics.  Significant 
theoretical effort has therefore been directed during the past several years 
towards determining the precise size of strong interaction corrections to
these weak transitions \refs{\Flynn{--}\Misiak}.  Progress has also
been made in estimating the hadronic matrix elements that characterize
the long distance aspects of these processes.  Chiral perturbation
theory calculations indicate that the CP conserving component of
$\kpiee$ decay is significantly smaller than the total CP violating
contribution, and the direct CP violating portion is believed to
dominate over its indirect counterpart \refs{\Cohen,\Ecker}.
\foot{This favorable hierarchy for CP conserving and violating contributions 
to $\kpiee$ decay has recently been challenged in ref.~\Donoghue.}
This kaon mode can thus provide an important window onto the nature of CP 
violation. 

	Much less is known about the rates at which the two
semileptonic FCNC reactions proceed within the MSSM.  The sensitivity
of $\kpiee$ to new supersymmetric physics has received little
attention.  We therefore examine the maximal variation in the MSSM
rate for the kaon process relative to its Standard Model value in this
article.  Supersymmetric contributions to $\Bsll$ decay were
previously considered by Bertolini {\it et al.} in ref.~\Bertolini.
These authors' conclusions need to be updated in light of the $\Bsg$
measurement.  The important constraint placed by the CLEO observation
upon the allowed MSSM parameter space was incorporated into the more
recent $\Bsll$ analysis of ref.~\Ali, but the impact of a universal
form for soft supersymmetry breaking terms was not studied in this 
latter work. As we shall see, inclusion of both the $\Bsg$ restriction
and the universality constraint disallows sizable deviations from the 
Standard Model in the integrated $\Bsll$ decay rate. Our work thus builds upon 
and extends previous supersymmetric FCNC investigations presented in the 
literature.

	Our paper is organized as follows.  In section~2, we review the 
elements of the MSSM which are relevant to our FCNC analysis.  We then 
discuss current restrictions upon the MSSM parameter space in section~3 and 
describe two different procedures for mapping out its allowed regions. 
We use these scanning algorithms to examine the impact of supersymmetry upon 
$\kpiee$ and $\Bsll$ decay in sections~4 and 5.  Finally, we close with a 
summary of our findings in section~6. 

% ==============================================================================
\newsec{The Minimal Supersymmetric Standard Model} 
% ==============================================================================

	The basic structure of the MSSM is well-known and has been thoroughly 
discussed in the literature \refs{\HaberKane{--}\Derendinger}.  We therefore 
recall just those aspects of the theory which are pertinent to $\kpiee$ and 
$\Bsll$ decay.  We first display our nomenclature conventions for matter 
superfields and their left handed fermion and scalar components in table~I.  
The fields listed in the first five rows of this table carry a generation 
subscript which ranges over three family values.  They are also assigned 
negative parities under a discrete $Z_2$ symmetry in order to forbid baryon 
and lepton number violating interactions.  

\vfill\eject

$$ \vbox{\offinterlineskip
\def\tablerule{\noalign{\hrule}}
\hrule
\halign {\vrule#& \strut#&
\ \hfil#\hfil & \vrule#&
\ \hfil#\hfil & \vrule#&
\ \hfil#\hfil & \vrule# \cr
\tablerule%
height10pt && \omit && \omit && \omit &\cr
&& Superfields && Fermions && Scalars &\cr
height10pt && \omit && \omit && \omit & \cr
\tablerule
height10pt && \omit && \omit && \omit &\cr
&& \quad $ Q_i=\pmatrix{U_i \cr D_i \cr}$ \quad && 
\quad $q_i = \pmatrix{u_i \cr d_i \cr}$ \quad && 
\quad $\tilde{q}_i = \pmatrix{\tilde{u}_i \cr \tilde{d}_i \cr}$ \quad &\cr
height10pt && \omit && \omit && \omit &\cr
&& \quad $U_i^c$ \quad && 
\quad $u_i^c$ \quad && \quad $\tilde{u}_i^c$ \quad &\cr
height10pt && \omit && \omit && \omit &\cr
&& \quad $D_i^c$ \quad && 
\quad $d_i^c$ \quad && \quad $\tilde{d}_i^c$ \quad &\cr
height10pt && \omit && \omit && \omit &\cr
&& \quad $L_i=\pmatrix{N_i \cr E_i \cr}$ \quad && 
\quad $\ell_i = \pmatrix{\nu_i \cr e_i \cr}$ \quad && 
\quad $\tilde{\ell}_i = \pmatrix{\tilde{\nu}_i \cr \tilde{e}_i \cr}$ \quad &\cr
height10pt && \omit && \omit && \omit &\cr
&& \quad $E_i^c$ \quad && 
\quad $e_i^c$ \quad && \quad $\tilde{e}_i^c$ \quad &\cr
height10pt && \omit && \omit && \omit &\cr
&& \quad $H_1=\pmatrix{H^0_1 \cr H^-_1 \cr}$ \quad && 
\quad $\tilde{h}_1 = \pmatrix{\tilde{h}_1^0 \cr \tilde{h}_1^- \cr}$ \quad && 
\quad $h_1 = \pmatrix{{h^0_1}^* \cr -h_1^- \cr}$ 
\quad &\cr
height10pt && \omit && \omit && \omit &\cr
&& \quad $H_2=\pmatrix{H^+_2 \cr H^0_2 \cr}$ \quad && 
\quad $\tilde{h}_2 = \pmatrix{\tilde{h}_2^+ \cr \tilde{h}_2^0 \cr}$ \quad && 
\quad $h_2 = \pmatrix{h_2^+ \cr h_2^0 \cr}$ 
\quad &\cr
height10pt && \omit && \omit && \omit &\cr
\tablerule}} $$
\centerline{Table I.  MSSM matter content}

\vskip 0.5 in

\noindent The Higgs fields appearing in the last two rows transform positively
under this matter parity symmetry.

	The chiral superfields in table~I enter into the superpotential  
\eqn\superpotential{W = \bigmu H_1 H_2 + Y^\U_{ij} Q_i U^c_j H_2
+ Y^\D_{ij} Q_i D^c_j H_1 + Y^\E_{ij} L_i E^c_j H_1}
which governs the supersymmetry preserving interactions among matter 
fields.  
\foot{Our sign convention for contracting two $SU(2)$ doublets is 
exemplified by the expansion $H_1 H_2 = H_1^0 H_2^0 - H_1^- H_2^+$ of the 
superpotential $\bigmu$ term.}
After vector superfield terms are included, the supersymmetric 
Lagrangian schematically appears in component form as 
\eqn\LSUSY{\eqalign{
\CL_{\rm SUSY} &= -\quarter {F_\G^\A}^{\u\v} {F_\G^\A}_{\u\v}
 + \bar{\lambda_\G^\A} i \Dslash_{\A\B} \lambda_\G^\B 
 + (D^\mu \phi)^\dagger (D_\mu \phi) + \bar{\psi} i \Dslash \psi \cr
 & - \left[ \left( {d W \over d \Phi_i} \right)^* 
     \left( {d W \over d \Phi_i} \right) + \half \left( 
     {\partial^2 W \over \partial \Phi_i \partial \Phi_j} 
     \psi_i^\T C \psi_j + \hc \right) \right]_{\Phi \to \phi} \cr
 & - \sqrt{2} g_\G \left[ \phi^\dagger T^\A_\G {\lambda^\A_\G}^\T C \psi 
     + \hc \right] - \half g^2_\G ( \phi^\dagger T^\A_\G \phi) 
        ( \phi^\dagger T^\A_\G \phi) .}}
The index $G$ labels the color, weak isospin and hypercharge factors in 
the Standard Model gauge group, and indices $A$ and $B$ range over the 
nonabelian subgroups' adjoint representations.  All MSSM scalars are 
assembled into $\phi$, while matter fermions and gauginos are 
respectively contained within the four-component left handed $\psi$ and 
$\lambda$ fields. 

	Since supersymmetry is manifestly violated in the low energy world, 
the MSSM Lagrangian is supplemented with the soft supersymmetry breaking terms
\eqn\Lsoft{\eqalign{
\CL_{\rm soft} = & -\half \bigl[ m_{\tilde{g}} \tilde{g}^{a \,\T} C \tilde{g}^a
+ m_{\tilde{\W}} \tilde{W}^{i \,\T} C \tilde{W}^i
+ m_{\tilde{\B}} \tilde{B}^\T C \tilde{B} + \hc \bigr]
- m_1^2 h_1^\dagger h_1 - m_2^2 h_2^\dagger h_2 \cr
& - \tilde{q}^\dagger_i (M^2_{\tilde{q}})_{ij} \tilde{q}_j 
- \tilde{u}^{c\,\dagger}_i (M^2_{\tilde{u}^c})_{ij} {\tilde{u}}^c_j 
- \tilde{d}^{c\,\dagger}_i (M^2_{\tilde{d}^c})_{ij} {\tilde{d}}^c_j 
- \tilde{\ell}^\dagger_i (M^2_{\tilde{\ell}})_{ij} \tilde{\ell}_j 
- \tilde{e}^{c\,\dagger}_i (M^2_{\tilde{e}^c})_{ij} {\tilde{e}}^c_j \cr
& 
+ \bigl[ A^\U_{ij} \tilde{q}_i \tilde{u}^c_j h_2 
+ A^\D_{ij} \tilde{q}_i \tilde{d}^c_j h_1
+ A^\E_{ij} \tilde{\ell}_i \tilde{e}^c_j h_1
+ B \bigmu h_1 h_2 + \hc \bigr]. \cr}}
In order to cut down the number of free parameters which enter into this 
expression to a manageable size, some relations among soft sector masses and 
couplings must be adopted.  We shall first assume that the weak scale
values of all the parameters in \Lsoft\ are simply related to GUT scale 
progenitors.  The running gaugino masses $m_{\tilde g}(\mu)$, 
$m_{\tildeW}(\mu)$ and $m_{\widetilde B}(\mu)$ then unify at $\mu=\MGUT$ just 
like the gauge couplings.  We also equate all scalar mass parameters with
a single $m_0$ at the GUT scale.  Finally, we set the trilinear interaction 
matrices $A^{\U,\D,\E}$ equal to $A_0 Y^{\U,\D,\E}$ at $\mu=\MGUT$ where $A_0$ 
denotes a common proportionality constant.  
\foot{We do not assume any {\it a priori} relationship between $A_0$ and $B$ 
in \Lsoft.}
Imposition of this universal structure upon soft supersymmetry breaking terms 
lends predictive power to the MSSM, but the assumed simplifications are quite 
strong.  We will therefore later relax some of these constraints and 
investigate a more general class of supersymmetric models.

	Renormalization group evolution of MSSM parameters down from the 
unification scale can generate a vacuum instability \refs{\Ibanez{--}\Wise}.
As the couplings in the scalar potential run, the neutral Higgs fields may at 
some point develop nonzero vacuum expectation values 
$\langle h_1^0 \rangle = v_1/\sqrt{2}$ and 
$\langle h_2^0 \rangle = v_2/\sqrt{2}$ which break electroweak symmetry.  The 
numerical value $v = \sqrt{v_1^2+v_2^2} = 246 \GeV$ for their mean is 
fixed by the $W$ boson mass.  But the ratio $\tan\beta = v_2/v_1$ 
remains a free parameter in the model.  We restrict this ratio 
to the range $2 \le \tan\beta \le 55$ so that Landau poles do not develop in 
the top or bottom Yukawa couplings anywhere between the weak and GUT scales. 

	Electroweak symmetry breaking induces mixing among MSSM fields. 
In the matter sector, primed mass eigenstates are related to unprimed 
gauge eigenstate counterparts as follows:
\eqn\masseigenstates{\eqalign{
u' &= S^\UL u + S^\UR C {\bar{u^c}}^{\>\T} \cr
d' &= S^\DL d - S^\DR C {\bar{d^c}}^{\>\T} \cr
\nu' &= S^\NL \nu \cr
e' &= S^\EL e - S^\ER C {\bar{e^c}}^{\>\T} \cr}
\qquad\qquad
\eqalign{
\tilde{u}' &= \Gamma^\U \pmatrix{S^\UL \> \tilde{u} \cr S^\UR \> 
\tilde{u^c}^*
\cr} \cr
\tilde{d}' &= \Gamma^\D \pmatrix{S^\DL \> \tilde{d} \cr -S^\DR \> 
\tilde{d^c}^*
\cr}\cr
\tilde{\nu}' &= \Gamma^\N S^\EL \tilde{\nu} \cr
\tilde{e}' &= \Gamma^\E \pmatrix{S^\EL \> \tilde{e} \cr -S^\ER \> 
\tilde{e^c}^*
\cr}. \cr}}
The unitary $S$ and $\Gamma$ transformations rotate fermion and sfermion mass 
matrices into real and diagonal forms.   The $3\times 3$ quark and lepton mass 
matrices are simply related to the Yukawa couplings in the superpotential:
\eqn\massmatrices{\eqalign{
M_\U &= {v \sin\b \over \sqrt{2}} S^\UR {Y^\U}^\T {S^\UL}^\dagger \cr
M_\D &= {v \cos\b \over \sqrt{2}} S^\DR {Y^\D}^\T {S^\DL}^\dagger \cr
M_\E &= {v \cos\b \over \sqrt{2}} S^\ER {Y^\E}^\T {S^\EL}^\dagger. \cr}}
On the other hand, the $6\times 6$ squared mass matrices for the squarks and 
sleptons look much more complicated and involve many parameters from both the 
supersymmetry conserving and violating Lagrangians in \LSUSY\ and \Lsoft:
$$ \eqalignno{
& M^2_{\tilde{u}} = \cr
& \Gamma^\U \pmatrix{ S^\UL M^2_{\tilde{q}} {S^\UL}^\dagger 
+ M_\U^2 + \displaystyle{\mZsq \over 6} (3-4\sin^2 \th) \cos 2\b 
& \bigmu M_\U \cot\b -\displaystyle{v \sin\beta \over \sqrt{2}} 
S^\UL {A^\U}^* {S^\UR}^\dagger \cr
\bigmu^* M_\U \cot\b -\displaystyle{v \sin\beta \over \sqrt{2}} 
S^\UR {A^\U}^\T {S^\UL}^\dagger &
S^\UR {M^2_{\tilde{u}^c}}^\T {S^\UR}^\dagger + M_\U^2 + \displaystyle{2 \mZsq 
\over 3} \sin^2\th \cos 2 \b} {\Gamma^\U}^\dagger \cr
& \cr
& M^2_{\tilde{d}} = \cr
& \Gamma^\D \pmatrix{ S^\DL M^2_{\tilde{q}} {S^\DL}^\dagger 
+ M_\D^2 - \displaystyle{\mZsq\over 6} (3-2\sin^2 \th) \cos 2\b & \bigmu M_\D 
\tan\b -\displaystyle{v \cos\beta \over \sqrt{2}} S^\DL {A^\D}^* 
{S^\DR}^\dagger \cr
\bigmu^* M_\D \tan\b -\displaystyle{v \cos\beta \over \sqrt{2}} 
S^\DR {A^\D}^\T {S^\DL}^\dagger &
S^\DR {M^2_{\tilde{d^c}}}^\T {S^\DR}^\dagger+M_\D^2 - \displaystyle{\mZsq\over 
3} \sin^2\th \cos 2 \b} {\Gamma^\D}^\dagger \cr} $$
\vfill\eject

\eqn\supermassmatrices{\eqalign{
& M^2_{\tilde{\nu}} = \Gamma^\N \bigl( S^\EL M^2_{\tilde{\ell}} {S^\EL}^\dagger 
+ \half \mZsq \cos 2\b \bigr) {\Gamma^\N}^\dagger \cr
& \cr
& M^2_{\tilde{e}} = \cr
& \Gamma^\E \pmatrix{ S^\EL M^2_{\tilde{\ell}} {S^\EL}^\dagger 
+ M_\E^2 - \displaystyle{\mZsq\over 2}(1-2\sin^2 \th) \cos 2\b 
& \bigmu M_\E \tan\b -\displaystyle{v \cos\beta \over \sqrt{2}} 
S^\EL {A^\E}^* {S^\ER}^\dagger \cr
\bigmu^* M_\E \tan\b -\displaystyle{v \cos\beta \over \sqrt{2}} 
S^\ER {A^\E}^\T {S^\EL}^\dagger &
S^\ER {M^2_{\tilde{e}^c}}^\T {S^\ER}^\dagger+M_\E^2 - \mZsq\sin^2\th 
\cos 2 \b} {\Gamma^\E}^\dagger. \cr}}

	Mixing also takes place in the gaugino and Higgs sectors.  The 
physical Dirac chargino and Majorana neutralino eigenstates are linear 
combinations of left handed Winos, Binos and Higgsinos: 
\eqn\gauginos{\eqalign{
\tilde{\chi}^- &= U \pmatrix{\tilde{W}^- \cr \tilde{h}_1^- \cr}
+ V^* C \pmatrix{\bar{\tilde{W}^+}^{\>\T} \cr 
                 \bar{\tilde{h}_2^+}^{\>\T} \cr} \cr
\tilde{\chi}^0_\M &= 
N \pmatrix{\tilde{B} \cr \tilde{W}_3 \cr \tilde{h}_1^0 \cr
\tilde{h}_2^0 \cr} + N^* C
\pmatrix{\bar{\tilde{B}}^{\>\T} \cr \bar{\tilde{W}_3}^{\>\T} \cr
\bar{\tilde{h}_1^0}^{\>\T} \cr \bar{\tilde{h}_2^0}^{\>\T} \cr}. \cr}}
The unitary transformations $U$, $V$ and $N$ diagonalize these fields' mass 
matrices
\eqn\charginomassmatrix{\eqalign{
M_{\tilde{\chi}^\pm} &= U^* \pmatrix{
m_{\tilde{\W}} & \sqrt{2} \mW \sin\beta \cr
\sqrt{2} \mW \cos\beta & -\bigmu \cr} V^\dagger \cr}}
and
\eqn\neutralinomassmatrix{\eqalign{
M_{\tilde{\chi}^0} &= N^* \pmatrix{
m_{\tilde{\B}} & 0 & -\mZ \sin\th \cos\b & \mZ \sin\th \sin\b \cr 
0 & m_{\tilde{\W}} & \mZ\cos\th \cos\b & -\mZ\cos\th \sin\b \cr
-\mZ \sin\th \cos\b & \mZ\cos\th \cos\b & 0 & \bigmu \cr 
\mZ \sin\th \sin\b & -\mZ\cos\th \sin\b & \bigmu & 0 \cr} N^\dagger . \cr}}
Similarly, charged scalar mass eigenstates are combinations of $h_1^\pm$ and 
$h_2^\pm$: 
\eqn\chargedscalars{\pmatrix{\pi^\pm \cr h^\pm \cr} = 
\pmatrix{\cos\b & \sin\b \cr
        -\sin\b & \cos\b \cr}
\pmatrix{h_1^\pm \cr h_2^\pm \cr}.}
The $\pi^\pm$ would-be Goldstone bosons are absorbed via the Higgs mechanism 
into the longitudinal components of the $W^\pm$ gauge fields.  But the 
remaining $h^\pm$ bosons represent genuine propagating scalar degrees of 
freedom whose tree level squared masses equal $\mhsq = \mWsq + m_1^2 + m_2^2 
+ 2 |\bigmu|^2$. 

	After the gauge eigenstate fields in the supersymmetric Lagrangian 
\LSUSY\ are rewritten in terms of their mass eigenstate counterparts,
\foot{We suppress primes on mass eigenstate fields from here on.}
it is straightforward to work out the interactions of gluinos,
charginos and neutralinos with quarks and squarks.  We list below the
resulting terms which participate at one-loop order in $d_i \to d_j
\ell^+ \ell^-$ decay:
\eqn\Lchi{\eqalign{
\CL_{\tilde{g},\tilde{\chi}} = & -\sqrt{2} g_3 \sum_{a=1}^8 
\bar{\tilde{g}^a_\M} \tilde{d}^\dagger ( \Gamma^\DL \Pminus - 
\Gamma^\DR \Pplus) T^a d \cr
& + \sum_{\I=1}^2 \bar{\tilde{\chi}}^{\, -}_\I 
\tilde{u}^\dagger ( X_\I^\UL \Pminus + X_\I^\UR \Pplus) d +
\sum_{\I = 1}^4 (\bar{\tilde{\chi}}^0_\M)_\I 
\tilde{d}^\dagger ( Z_\I^\DL \Pminus + Z_\I^\DR \Pplus) d + \hc \cr}}
where
\eqn\XandZ{\eqalign{
X_\I^\UL &= g_2 \bigl[ -V^*_{\I1} \Gamma^\UL + V^*_{\I2} \Gamma^\UR
{M_\U \over \sqrt{2} \mW \sin\beta} \bigr] K \cr
X_\I^\UR &= g_2 \bigl[ U_{\I2} \Gamma^\UL K 
{M_\D \over \sqrt{2} \mW \cos\beta} \bigr] \cr
Z_\I^\DL &= -{g_2 \over \sqrt{2}} \bigl[ ( -N^*_{\I2} + \third \tan\th 
N^*_{\I1}) \Gamma^\DL +N^*_{\I3} \Gamma^\DR {M_\D \over \mW \cos\beta} \bigr] 
\cr
Z_\I^\DR &= -{g_2 \over \sqrt{2}} \bigl[ \twothirds \tan\th N_{\I1} \Gamma^\DR
+N_{\I3} \Gamma^\DL {M_\D \over \mW \cos\beta} \bigr].  \cr}}
Flavor mixing enters into these interactions through the Kobayashi-Maskawa 
matrix $K = S^\UL {S^\DL}^\dagger$ and the $6\times 3$ block components of 
$\Gamma^\U$ and $\Gamma^\D$:
\eqn\Gammamatrices{\eqalign{
\Gamma^\U_{6 \times 6} &= \pmatrix{\Gamma^\UL_{6 \times 3} & 
  \Gamma^\UR_{6 \times 3} \cr} \cr
\Gamma^\D_{6 \times 6} &= \pmatrix{\Gamma^\DL_{6 \times 3} & 
  \Gamma^\DR_{6 \times 3} \cr}. \cr}}
Other gauge boson and Higgs terms which mediate the FCNC processes of interest 
are similarly extracted from the Lagrangian.  The Feynman rules for all these 
interactions may be found in the literature \refs{\HaberKane,\Rosiek}.

	Having set up the basic MSSM framework, we are now ready to explore 
its large parameter space.  We take up this topic in the following section.

% ==============================================================================
\newsec{MSSM parameter space} 
% ==============================================================================

	Before predictions can be derived from the Minimal Supersymmetric 
Standard Model, explicit values for the parameters in the superpotential 
\superpotential\ and soft supersymmetry breaking Lagrangian \Lsoft\ must be 
specified.  In order to reduce the size of the parameter space, we initially 
adopt the common assumption that MSSM masses and couplings are simply related 
at $\mu=\MGUT$.  This ansatz is motivated by the simplest supergravity theories 
\SUGRA. A universal soft supersymmetry breaking sector at $\mu=\MGUT$ also
decreases the likelihood of generating unacceptably large FCNC
amplitudes at the weak scale.  Instead, the magnitudes for such
amplitudes in the MSSM are anticipated to be of the same order as
those in the Standard Model.  Experimentally discriminating between
the two theories' predictions for various FCNC transitions rates will
not be an easy task.  But this goal is hoped to be achieved by a
number of experimental programs within the next several years.

	In our analysis, we take as input parameters the dimensionful soft 
sector quantities $A_0(\MGUT)$, $m_0(\MGUT)$ and $m_{\tildeW}(\mZ)$, the 
dimensionless ratio $\tan\beta$, and all Standard Model fermion and gauge 
boson masses and couplings.  We also restrict the source of CP violation in 
the MSSM to stem from just a single phase in the KM matrix.  Imaginary parts of 
$m_{\tildeW}$ and $B \bigmu$ can be rotated away by field redefinitions, but 
$A_0$ and $\bigmu$ generally remain complex.  The phases of these last two 
parameters are tightly constrained by neutron electric dipole moment 
limits \Buchmuller.  We shall simply take $A_0$ and $\bigmu$ to be real.

	In order to determine the numerical values for all the
couplings in the MSSM, we follow a lengthy yet straightforward
procedure.  We first locate the GUT scale $\MGUT \sim 10^{16} \GeV$ by
evolving the $SU(2)_\L$ and $U(1)_\Y$ gauge couplings up to the point
where they meet.  We then choose specific values for $A_0(\MGUT)$,
$m_0(\MGUT)$, $m_{\tildeW}(\mZ)$ and $\tan\beta$.  A large value for
the trilinear scalar coupling in conjunction with small values for the
common scalar and gaugino mass parameters tends to yield stop masses
which are too light to satisfy direct search constraints.  Large
values for either $m_0$ or $m_{\tildeW}$ lead to squark decoupling and
negligible supersymmetric contributions to FCNC decays.  We therefore
restrict the magnitudes of the three dimensionful quantities to be
less than 1 TeV.

	After the gaugino, scalar and Yukawa terms in the soft supersymmetry 
breaking sector are evaluated at $\mu=\MGUT$ and run down to 
$\mu=\mZ$, the numerical values for all MSSM parameters except $\bigmu$ and $B$ 
are determined.  The tree level relations
\eqn\muandB{\eqalign{
|\bigmu|^2 &= {m_2^2 \sin^2\b - m_1^2 \cos^2\b \over \cos 2 \b} 
  	    - \half \mZ^2 \cr
B \bigmu &= \half \sin 2 \b(m_1^2 + m_2^2 + 2 |\bigmu|^2) \cr}}
fix these last two quantities up to a twofold ambiguity in $\rm{sgn(\bigmu)}$.  
Points in the MSSM parameter space which yield negative values for $|\bigmu|^2$ 
or $B \bigmu$ fail to break electroweak symmetry and are rejected.  Necessary 
tree level conditions for the existence of a stable scalar potential minimum
\eqn\stability{\eqalign{
& |B \bigmu|^2 > (m_1^2 + |\bigmu|^2)(m_2^2 + |\bigmu|^2) \cr
& (m_1^2-m_2^2) \cos 2 \b > 0 \cr}}
must also be satisfied.  We further require that the MSSM particle
spectrum be consistent with present limits from direct superpartner
searches \PDB.  In particular, we impose the recent LEP~1.5 lower bound of 
$65 \GeV$ on the chargino mass \LEPx.

	The final constraint which we place upon the MSSM parameter space 
comes from $\Bsg$ decay.  Supersymmetry modifies the Standard Model prediction 
for the rare radiative rate by adding extra contributions to the Wilson 
coefficients $C_7(\mW)$ and $C_8(\mW)$ of the electromagnetic and 
chromomagnetic moment operators in the $\Delta B = 1$ effective Hamiltonian 
\BMMP.  We neglect $C_8$ since it accounts for only 3\% of the Standard Model 
$b \to s \gamma$ amplitude and is not expected to be significantly more 
important in the MSSM.  But the charged Higgs, chargino, neutralino and gluino 
contributions to $C_7$ which are tabulated in Appendix~A can be quite 
substantial.  We therefore calculate the ratio
\eqn\Rseven{R_7 = {C_7(\mW)_\MSSM \over C_7(\mW)_\SM}}
at each point in the MSSM parameter space.  We throw away all points whose 
values for $R_7$ do not lie within the allowed intervals
\eqn\Rsevenlimits{0.4 < R_7 < 1.2 \qquad {\rm or} \qquad
		 -4.2 < R_7 < -2.4}
that take into account current experimental errors \CLEO\ and theoretical 
uncertainties \BMMP. 

	After scanning over the MSSM parameter space and imposing all 
the above criteria, we identify a number of general features which hold 
everywhere in the allowed regions except in the very large $\tan\b$ domain:
\item{(i)}  Sizable supersymmetry contributions to $\kpiee$, $\Bsll$ and 
$\Bsg$ decay mainly arise from charged Higgs and chargino exchange. 

\item{(ii)}  Flavor violating entries in $\Gamma^\U$ hardly affect these FCNC
processes.

\item{(iii)}  The first two generations of up and down squarks are almost 
degenerate.  The first two generations of sleptons and sneutrinos 
are also nearly degenerate.

\item{(iv)}  Left-right squark and slepton mixing is negligible for the first 
two generations.
\medskip\noindent
Similar observations have previously been noted in ref.~\Bertolini.

	These characteristics provide useful guidelines for establishing less 
restrictive constraints on supersymmetric extensions of the Standard Model 
than those which underlie the MSSM with a universal soft breaking sector.  
Rather than starting with a unified set of GUT scale parameters and evolving 
them down to low energies, we can instead survey all possible values for MSSM 
couplings and masses at the weak scale for which conditions (i) - (iv) are 
satisfied.  This alternate mapping procedure provides a useful check on the 
sensitivity of FCNC results upon the assumed form of the soft supersymmetry 
breaking sector.  The particular weak scale quantities which must be specified 
in order to determine supersymmetric contributions to 
$d_i \to d_j \ell^+ \ell^-$ decay are listed below:
\item{$\bullet$} a common mass $m_{\tilde{u}_\L}$ for the superpartners of left 
handed up and charm quarks, 
\item{$\bullet$} the masses $m_{\tilde{t}_\L}$, $m_{\tilde{t}_\R}$ and 
mixing angle $\at$ for top squarks,
\item{$\bullet$} a common mass $m_{\tilde{\nu}}$ for the first two generation 
sneutrinos, 
\item{$\bullet$} the Wino and charged Higgs masses $m_{\tildeW}$ and 
$m_{h^\pm}$,
\item{$\bullet$} the superpotential $\bigmu$ parameter and $\tan\beta$.
\smallskip\noindent
This parameterization is similar in spirit to the one adopted in ref.~\Ali.
As in our universal soft sector analysis, we shall restrict the dimensionful 
quantities to the sub-TeV regime and restrict the dimensionless VEV ratio to 
$2 \le \tan\beta \le 55$. 

	In the next two sections, we will investigate the $\kpiee$ and $\Bsll$ 
transitions utilizing the universal soft sector scanning procedure as well as 
the more general mapping method.  The greater labor required to explore 
the much larger nine-dimensional parameter space in the latter approach 
is offset by several simplifications.  For example, neutralino and gluino 
contributions to $d_i \to d_j \ell^+ \ell^-$ may simply be neglected.  The 
$\Gamma^\U$ matrix also reduces to the nearly diagonal form 
\eqn\newGamma{\Gamma^\U = \pmatrix{1 & 0 & 0 & 0 & 0 & 0 \cr
				   0 & 1 & 0 & 0 & 0 & 0 \cr
				 0 & 0 & \cos\at & 0 & 0 & -\sin\at \cr
				   0 & 0 & 0 & 1 & 0 & 0 \cr
				   0 & 0 & 0 & 0 & 1 & 0 \cr
				 0 & 0 & \sin\at & 0 & 0 & \cos\at \cr}}
when criteria (i) - (iv) are imposed.  But most importantly, no renormalization 
group evolution needs to be performed.  Searching the nine-dimensional 
parameter space for sets of points where supersymmetry significantly enhances 
or suppresses the rare decay modes is thus rendered tractable.

% ==============================================================================
\newsec{$\pmb{\kpiee}$ decay}
% ==============================================================================

	The total amplitude for $\kpiee$ decay can be decomposed into CP 
conserving and violating parts.  The former starts at second order in the 
electromagnetic interaction.  As a result, the CP conserving branching 
fraction $\Br(\kpiee)_{CP} \simeq (0.3 \dash 1.8) \times 10^{-12}$ is 
significantly smaller than its CP violating counterpart \Cohen.  Moreover, the 
indirect component of the CP violating amplitude is believed to be smaller than 
the direct part.  Present data imply $\Br(\kpiee)_{\rm indirect} \le 1.6 
\times 10^{-12}$ \Ecker, while the Standard Model prediction for the direct CP 
violating contribution lies in the range $\Br(\kpiee)_{\rm direct} \simeq 
(2.5 \dash 9.0) \times 10^{-12}$ \Buras.
\foot{The sizable uncertainty in the direct CP violating branching fraction 
primarily stems from KM angles which are poorly constrained at present.  The 
predictions' precision should substantially improve when the KM unitarity 
triangle is better determined in upcoming B-factory studies.}
The CP conserving and indirect CP violating amplitudes may be computed using 
experimentally determined values for chiral Lagrangian coefficients and the 
$\epsilon$ parameter which automatically include possible MSSM contributions. 
Since these quantities are consistent with Standard Model predictions, we will 
focus exclusively upon the direct CP violating component in our supersymmetric 
FCNC investigation. 
	
	The analysis of $\kpiee$ decay is greatly facilitated by working 
within an effective field theory framework.  In this approach, heavy degrees 
of freedom are successively integrated out from a specified full theory, 
and the resulting effective theory is run down to a low energy hadronic scale 
using the renormalization group.  A finite number of flavor changing 
dimension-6 operators generated by this process at $\mu=\mW$ enter into the 
effective Hamiltonian $\Heff$ which governs the dynamics of the low energy 
theory.  The complete set of left handed $\Delta S=1$ four-fermion operators 
that originate from the Standard Model and mediate $\kpiee$ are catalogued in 
ref.~\Buras.  If the starting full theory is taken to be the MSSM, additional 
terms with right handed flavor changing currents appear in $\Heff$.  But since 
their coefficients are tiny, the extra operators may be neglected without 
loss.  

	After the effective theory is evolved to low energies, the direct CP 
violating $\kpiee$ amplitude is well approximated by the long distance matrix 
element $\langle \pi^0 e^+ e^- | \CH_{\rm eff} | K_L \rangle$ of the truncated 
Hamiltonian
\eqn\Heffective{\Heff = {-\GF \over \sqrt{2}} \KMd
\bigl[ y_{7V} Q_{7V} + y_{7A} Q_{7A} \bigr]  +  \hc}
All information associated with short distance physics is encoded into the 
Wilson coefficients
\eqn\coefficients{\eqalign{
y_{7V} &= {\aEM \over 2 \pi \sinsq} \Bigl[ P_0 + \bigl(Y(\xt)+Y^\SUSY 
\bigr) - 4 \sinsq \bigl(Z(\xt)+Z^\SUSY \bigr) + P_E \bigl( E(\xt) + 
E^\SUSY \bigr) \Bigr] \cr
y_{7A} &= - {\aEM \over 2 \pi \sinsq} \Bigl[Y(\xt)+Y^\SUSY \Bigr] \cr}}
of the semileptonic operators 
\eqn\Qoperators{\eqalign{
Q_{7V} &= \sbar \gamma^\mu (1-\gfive) d \> \ebar \gamma_\mu e \cr
Q_{7A} &= \sbar \gamma^\mu (1-\gfive) d \> \ebar\gamma_\mu \gfive e. \cr}}
The functions 
\eqn\SMmatching{\eqalign{
Y(x_t) &= {4x_t-x_t^2 \over 8 (1-x_t)} + {3x_t^2 \over 8(1-x_t)^2} \log x_t \cr
Z(x_t) &= {108x_t - 259 x_t^2 + 163 x_t^3 - 18 x_t^4 \over 144(1-x_t)^3} 
	+ {-8 + 50 x_t - 63 x_t^2 - 6 x_t^3 + 24 x_t^4 \over 72 (1-x_t)^4} 
  \log x_t \cr 
E(x_t) &= {18 x_t - 11 x_t^2 - x_t^3 \over 12 (1-x_t)^3} - {4 - 16 x_t + 
9 x_t^2 \over 6 (1-x_t)^4} \log x_t \cr}}
of the variable $x_t = (\mt/\mW)^2$ summarize the high energy contributions 
to $\kpiee$ decay that are common to both the Standard Model and MSSM.  
Next-to-leading order QCD corrections which are the same in both theories are 
incorporated into $P_0 \simeq 0.7$ and $P_E \simeq -0.01$ \Buras.  Therefore, 
all effects of supersymmetry upon the rare weak transition reside within the 
$Y^\SUSY$, $Z^\SUSY$ and $E^\SUSY$ parameters in eqn.~\coefficients.

	The values of the $y_{7V}$ and $y_{7A}$ coefficients are obtained 
after matching renormalized $\sbar \to \dbar e^+ e^-$ amplitudes calculated in 
both the full and effective theories.  We perform this matching at $\mu=\mW$ 
in the Standard Model as well as in its minimal supersymmetric extension.  The 
one-loop bubble, penguin and box graphs which must be evaluated on the 
MSSM side of the matching condition are displayed in figs.~1a, 1b and 1c.  
After a long but straightforward computation, we find the charged Higgs, 
chargino, neutralino and gluino contributions listed in Appendix~B to 
the $Y^\SUSY$ and $Z^\SUSY$ terms in \coefficients.

	Several points about our supersymmetric matching results should be 
noted.  Firstly, we have not included any bubble or penguin graphs which 
involve neutral Higgs boson exchange in fig.~1a or fig.~1b.  Such diagrams 
are proportional to the electron mass $m_e$ and are negligibly small.  
Similarly, the first box graph in fig.~1c with a charged Higgs running around 
the loop vanishes in the $m_e \to 0$ limit and may be safely ignored.
Secondly, we have intentionally not calculated $E^\SUSY$ which should be 
comparable in size to $E(\xt) \simeq 0.25$.  Since the numerical value for 
$P_E$ is almost two orders of magnitude smaller than that for $P_0$, the term 
proportional to $P_E$ in \coefficients\ is negligible and $E^\SUSY$ is 
unimportant.  Thirdly, we have ignored gluino loop matching contributions to 
charged current four-quark operators in the $\Delta S=1$ effective 
Hamiltonian.  Such contributions affect $y_{7V}$ and $y_{7A}$ at the 
sub-percent level.  Finally, the matching condition results displayed in 
Appendix~B are independent of any choice for the structure of soft 
supersymmetry breaking terms.

	The short distance $y_{7V}$ and $y_{7A}$ Wilson coefficients enter 
into the direct CP violating $\kpiee$ partial width along with a long distance 
hadronic matrix element.  After neglecting tiny imaginary components in the 
former quantities and relating the latter to the measured 
$K^+ \to \pi^0 e^+ \nu$ rate via an isospin rotation, we find \Buras
\eqn\kpieerate{\Br(\kpiee)_{\rm direct}= \Br(K^+ \to \pi^0 e^+ \nu) {\tau(K_\L) 
\over \tau(K^+)} \left | { \Im(K^*_{ts} K_{td}) \over K_{us} } \right |^2 
(y_{7V}^2 + y_{7A}^2).}
The sensitive dependence of this expression upon poorly known KM angles can 
be removed by normalizing the MSSM branching fraction to its Standard Model 
analogue.  The ratio
\eqn\kpieeratio{\rhokpiee = {\Br(\kpiee)_\MSSM - \Br(\kpiee)_\SM \over 
\Br(\kpiee)_\SM}}
thus cleanly quantifies supersymmetric enhancement or suppression of the rare 
kaon mode's rate.

	Following the universal soft sector scanning procedure outlined in 
section~3, we evaluate $\rhokpiee$ in the allowed regions of MSSM parameter 
space.  Representative results for two-dimensional slices through this space 
are illustrated in fig.~2a and fig.~2b.  In the first LEGO block figure, 
$\rhokpiee$ is plotted as a function of $A_0(\MGUT)$ and $m_0(\MGUT)$ 
with $m_{\tildeW}(\mZ) = 90 \GeV$ and $\tan\beta=5.0$ held fixed.  In the 
second figure, we display $\rhokpiee$ as a function of $m_{\tildeW}(\mZ)$ 
and $\tan\beta$ with $A_0(\MGUT)=-500\GeV$ and $m_0(\MGUT)=100\GeV$.  We have 
taken the sign of the $\bigmu$ parameter in the superpotential to be positive 
in both plots.  Points in the MSSM parameter space that are excluded by one or 
more of our imposed criteria are indicated in these figures by LEGO blocks 
which are saturated at their maximum values.  Looking at the results in 
figs.~2a and 2b, we see that supersymmetric effects in the MSSM with a 
universal soft sector reduce the direct CP violating $\kpiee$ branching 
fraction relative to the expected Standard Model rate by at most 10\%.   
These discrepancies between the Standard Model and its minimal supersymmetric 
extension are unfortunately too small to be realistically detected by 
upcoming experiments within the next several years.  

	It is interesting to examine whether larger deviations could result 
if some of the stringent assumptions underlying the universal soft sector 
MSSM are relaxed.  We have therefore performed several scans involving 
hundreds of thousands of points over the more general nine-dimensional 
parameter space discussed in section~3.  These scans reveal that pockets in 
the larger space exist where discrepancies between the Standard Model and the 
MSSM are three times larger in both the positive and negative directions than 
those we previously uncovered in our more restricted searches.  But even a 
30\% supersymmetric enhancement or suppression of the $\kpiee$ rate relative 
to its Standard Model value is unlikely to be experimentally resolvable in 
the near future. 

	The potential impact of supersymmetry upon the rare kaon mode is thus 
disappointingly small.  But as we shall see in the next section, the 
prospects for detecting signs of supersymmetry in $\Bsll$ decay are brighter.

% ==============================================================================
\newsec{$\pmb{\Bsee}$ and $\pmb{\Bsmm}$ decay}
% ==============================================================================

	Inclusive $\Bsee$ and $\Bsmm$ decay share several similarities with 
the $\kpiee$ transition.  Like their kaon analogue, these $B$ meson 
reactions are most conveniently analyzed within a low energy effective theory 
framework.  The complete list of dimension-6 operators in the $\Delta B=1$ 
effective Hamiltonian that participate in $\Bsll$ decay may be found in 
ref.~\GSW.  The Wilson coefficients of the bottom sector operators evaluated 
at the $W$ scale are trivially related to their strange sector analogues.  The 
next-to-leading order evolution of these coefficients from $\mu = \mW$ to 
$\mu = m_b$ has been calculated in ref.~\Misiak.  The details of this strong 
interaction running are quite complicated, and we will not present them here.  
Instead, we will simply apply the main results in our study of supersymmetric 
effects upon the rare decay modes. 

	The inclusive rate for the meson level process $\Bsll$ may be 
approximated by the rate for the free quark transition $b \to s \ell^+ \ell^-$ 
\Falk. Two independent variables are required to describe the latter process. 
We choose them to be the rescaled lepton energies
\eqn\leptonenergies{y_+ = {2 E_{\ell^+} \over m_b} \qquad {\rm and} \qquad 
	     y_- = {2 E_{\ell^-} \over m_b} }
measured in the $b$ quark rest frame.  When expressed in terms of these 
variables along with the rescaled squared invariant mass $\shat = y_+ + y_- -1$ 
of the lepton pair, the differential decay rate looks like 
\eqn\diffrate{\eqalign{&
{d^2 \Gamma (b \to s \ell^+ \ell^-) \over dy_+\;dy_-} = 
{\GF^2 m_b^5 |K_{ts}^* K_{tb}|^2 \over 16 \pi^3} 
\Bigl({ \aEM \over 4 \pi} \Bigr)^2 \Biggl\{ 
\Bigl[y_+(1-y_+)+y_-(1-y_-)\Bigr]\bigl( |C_9^\eff (\shat)|^2 + 
C_{10}^2 \bigr) \cr
& \qquad\quad + {4 \over \shat} 
\Bigl[ \shat(1 -\shat) + (1-y_+)^2 + (1-y_-)^2 + {2 m_\ell^2 \over \shat m_b^2}
\Bigr] \;(C_7^\eff)^2 \cr
& \qquad\quad + 4 ( 1- \shat ) C_7^\eff {\rm Re}(C_9^\eff(\shat)) 
+ 2(y_+ - y_-)C_{10}\; 
\Bigl[2 C_7^\eff + \shat {\rm Re}(C_9^\eff(\shat)) \Bigr]
\Biggr\}. \cr}}
The quantities 
\eqna\effcoeffs
$$ \eqalignno{
C_7^\eff &= C_7(\mW) \eta^{16/23}+{8 \over 3} C_8(\mW) 
\bigl( \eta^{14/23} - \eta^{16/23} \bigr) + \sum_{i=1}^8 h_i \eta^{a_i} & 
\effcoeffs a \cr
C_9^\eff &=\left( {\pi \over \aS(\mW)}+{\omega(\shat)\over\eta}\right)
\bigl( -0.1875 + \sum_{i=1}^8 p_i \eta^{a_i+1} \bigr) & \cr
& \qquad + {Y(x_t) + Y^\SUSY \over \sin^2\theta} - 4(Z(x_t) + Z^\SUSY) 
    + (E(x_t) + E^\SUSY) \bigl( 0.1405 + \sum_{i=1}^8 q_i \eta^{a_i+1} \bigr) 
& \cr
& \qquad + 1.2468 + \sum_{i=1}^8 \eta^{a_i} \Bigl[ r_i + s_i \eta + t_i 
h({\mc \over \mb},\shat) + u_i h(1,\shat) + v_i h(0,\shat) \Bigr] & 
\effcoeffs b \cr
C_{10} &= -{Y(x_t) + Y^\SUSY \over \sin^2\theta} & \effcoeffs c \cr } $$
that enter into the partial width depend upon the strong interaction coupling
ratio $\eta = \aS(m_{\W})/\aS(m_b)$ and various matching condition functions 
which we previously encountered in our $\kpiee$ analysis.  The effective 
coefficients also involve the components of the following 8-dimensional 
vectors:
\eqn\array{\eqalign{
a_i &= ( 0.6087, 0.6957, 0.2609, -0.5217,   
        0.4086, -0.4230, -0.8994, 0.1456), \cr
h_i &= ( 2.2996, -1.0880, -0.4286, -0.0714,
        -0.6494, -0.0380, -0.0186, -0.0057), \cr
p_i &= ( 0,      0,      -0.3941, 0.2424,
        0.0433, 0.1384, 0.1648, -0.0073), \cr
q_i &= ( 0,      0,      0,      0,
       0.0318, 0.0918, -0.2700, 0.0059), \cr
r_i &= ( 0,      0,      0.8331, -0.1219,       
	-0.1642, 0.0793, -0.0451, -0.1638), \cr
s_i &= ( 0,      0,      -0.2009, -0.3579,
       0.0490, -0.3616, -0.3554, 0.0072), \cr
t_i &= ( 0,      0,      1.7143, -0.6667,
       0.1658, -0.2407, -0.0717, 0.0990), \cr
u_i &= ( 0,      0,      0.2857, 0, 
       -0.2559, 0.0083,   0.0180, -0.0562), \cr
v_i &= ( 0,      0,      0.1429, 0.1667,
        -0.1731, -0.1120, -0.0178, -0.0067). \cr}}
Finally, the functions $h(z,\shat)$ and $\omega(\shat)$ which appear in 
eqn.~\effcoeffs{b}\ are given by 
\eqn\hwfuncs{\eqalign{
h(z,\shat) &= -{8 \over 9} \log z + {8 \over 27} + {4 \over 9} x 
-{2 \over 9}(2+x)\sqrt{|1-x|} 
\cases{\log \left| {\sqrt{1-x}+1 \over \sqrt{1-x}-1}\right|-i\pi, 
&  for $x \equiv 4z^2/\shat < 1$ \cr
2\;{\rm arctan}(1/\sqrt{x-1}), & for $x \equiv 4z^2/\shat > 1$ \cr} \cr
\omega(\shat) &= -{4 \over 3} \Li_2(\shat) -{2 \over 3} \log(\shat) 
\log(1-\shat) - {2 \over 9} \pi^2
-{5+4\shat \over 3(1+2\shat)} \log (1-\shat) \cr
&\qquad -{2\shat(1+\shat)(1-2\shat) \over 3(1-\shat)^2(1+2\shat)}
\log(\shat)+{5+9\shat-6\shat^2 \over 6(1-\shat)(1+2\shat)}.\cr }}
In order to consistently compute the differential rate to one-loop order 
accuracy, we only retain terms in $d^2 \Gamma / dy_+ dy_-$ up to linear order 
in $\omega(\shat)$.   We also set $\omega(\shat)$ to zero in the interference 
terms in eqn.~\diffrate\ which are proportional to $\Re(C_9^\eff(\shat))$.

	The Standard Model prediction for the $\Bsll$ decay rate is 
simply recovered from these formulae by setting $Y^\SUSY$, $Z^\SUSY$ and 
$E^\SUSY$ to zero and equating $C_7(m_{\W})$ and $C_8(m_{\W})$ with their 
Standard Model values.   In this case, the partially integrated rate
$d\Gamma/d\shat$ differs from the corresponding result discussed in 
ref.~\Misiak\ by a term proportional to the lepton mass.  This term was 
previously neglected because its effect is very small over nearly all of 
phase space.  However, its contribution to the integrated decay rate is not 
suppressed by $m_\ell$ provided no lower cut on $\shat$ is imposed.
As no such cut has been performed in the recent CDF analysis of $B \to K^* 
\mu^+ \mu^-$ \CDF, we retain this additional term in $d\Gamma/d\shat$. 

	The matching condition expressions for $Y^\SUSY$ and $Z^\SUSY$ that 
enter into the differential rate for $\Bsee$ trivially differ from those for 
$\kpiee$ by just the flavor label renamings specified in Appendix~B.  As in 
our previous $\kpiee$ matching computation, we neglect $E^\SUSY$ and set 
$C_8(\mW)$ to zero.  For the $\Bsmm$ mode, slepton and sneutrino indices must 
also be transformed from the first to second generation.  The numerical values 
for individual matching contributions to $Y^\SUSY$ and $Z^\SUSY$ turn out to 
be almost identical for all three FCNC channels which we consider in this 
paper.  The flavor independence of $Y^\SUSY$ and $Z^\SUSY$ implies that 
regions of the $\Bsll$ spectrum dominated by $C_9^\eff$ and $C_{10}$ are no 
more sensitive to supersymmetry than $\kpiee$ decay.  Sizable discrepancies 
between the Standard Model and its minimal supersymmetric extension can 
therefore only arise in $\Bsll$ observables which depend to a large extent 
upon $C_7^\eff$.

	We should note an interesting point regarding the supersymmetric limit 
of these matching results.  The approximate cancellation between different 
superpartner contributions to $\Bsg$ decay has been interpreted as a 
manifestation of $\Gamma(b \to s \gamma)=0$ in the supersymmetric limit \BG.  
Since the magnetic moment operator which mediates the radiative transition 
belongs to a linear multiplet, it cannot arise in a fully supersymmetric 
effective Hamiltonian \Ferrara.  No analogous argument can be made for 
$\bsll$ decay.  Penguin and box diagrams generate effective four-fermion
operators that form the highest components of vector superfields.  Such 
$D$-terms survive in the limit of exact supersymmetry. 

	As low statistics will most likely hinder experimental determination 
of the full differential spectrum, we need to consider various integrated 
observables.  Following the CDF analysis presented in ref.~\CDF, we first 
integrate the rate over the lepton pair mass regions 
\eqn\leppairinterval{m_{\ell^+ \ell^-} \in (2 m_\ell, 2.9\;\GeV) \cup 
(3.3\GeV, 3.6\GeV) \cup (3.8\GeV, 4.6\GeV).}
These disjoint intervals exclude $m_{\ell^+ \ell^-}$ values for which the 
inclusive $\Bsll$ rate is dominated by intermediate $\Jpsi$ and $\psi'$ 
states.  By restricting our analysis to just this nonresonant region,
we ensure the validity of the free quark approximation to inclusive $B$ meson 
decay.

	It is customary to reduce the uncertainties in the $\bsll$ 
partial width by normalizing it to the semileptonic rate
\eqn\semileptonicrate{\Gamma(b \to c e^+ \nu) =
{\GF^2 m_b^5 |K_{cb}|^2 \over 192 \pi^3} g\bigl({\mc \over \mb}\bigr) 
\left\{ 1 - {2 \aS(m_b) \over 3 \pi} \left[(\pi^2-{31 \over 4})
(1-{\mc \over \mb})^2 + {3 \over 2} \right] \right\} } 
which is related to the measured branching ratio 
$\Br(B \to X_c e^+ \nu) = 0.104 \pm 0.004$ \PDB.  The function 
\eqn\phasespace{
g(z) = 1 - 8z^2 + 8z^6 - z^8 - 24z^4\;{\rm log}\;z} 
appearing in this rate formula represents a phase space suppression factor.  
The sensitive dependence of $\Gamma(\bsll)$ and $\Gamma(b \to c e^+ \nu)$ upon 
KM angles cancels in their ratio.  However, errors in the numerical evaluation 
of the $\Bsll$ partial width can be reduced to only the $10 \dash 20 \%$ range 
due to uncertainties in quark masses and interference effects from excited 
charmonium states \Ligeti.  Therefore, signals of new physics beyond the 
Standard Model will be detectable only if they significantly exceed this 
level.

	After integrating the differential rate in eqn.~\diffrate, 
expressing the result in terms of the ratios
\eqn\Rvalues{
R_7 = {C_7(\mW)_{MSSM} \over C_7(\mW)_{SM}}, \qquad
R_Y = {Y(x_t)+Y^\SUSY \over Y(x_t)} \qquad {\rm and} \qquad 
R_Z = {Z(x_t)+Z^\SUSY \over Z(x_t)},}
and adopting the parameter values $\mc=1.3 \GeV$, $\mb=4.7 \GeV$, $\mt = 176 
\GeV$, $\aS(\mZ)=0.118$ and $|K_{ts}^* K_{tb} / K_{cb}|^2 = 0.95$, we find the 
following nonresonant branching fractions: 
\eqn\eebr{\eqalign{
\Br(B & \to X_s e^+ e^-)_{\rm NR} = 3.0 \times 10^{-7} 
[5.5 + 2.3 R_7^2 + 17.6 R_Y^2 + 3.7 R_Z^2  - 2.1 R_7 R_Y \cr 
& + 1.4 R_7 R_Z - 11.5 R_Y R_Z  + 4.6 R_7 + 8.1 R_Y - 5.3 R_Z]
\cr}}
\eqn\mumubr{\eqalign{
\Br(B &\to X_s \mu^+ \mu^-)_{\rm NR} = 3.0 \times 10^{-7} 
[2.9 + 0.8 R_7^2 + 17.5 R_Y^2 + 3.7 R_Z^2  - 2.1 R_7 R_Y \cr
& +1.4 R_7 R_Z - 11.4 R_Y R_Z  + 0.7 R_7 + 8.1 R_Y - 5.3 R_Z].
\cr}}
The Standard Model values $7.3 \times 10^{-6}$ and $4.9 \times 10^{-6}$ 
for the $\Bsee$ and $\Bsmm$ branching ratios are recovered by setting 
$R_7=R_Y=R_Z=1$ in these formulae.   

	The second $\bsll$ observable we consider is the lepton-antilepton 
energy asymmetry 
\eqn\Asymmetrydefn{
\CA = { N(E_{\ell^-} > E_{\ell^+}) - N(E_{\ell^+} > E_{\ell^-}) \over
          N(E_{\ell^-} > E_{\ell^+}) + N(E_{\ell^+} > E_{\ell^-})}.  }
Here $N(E_{\ell^-} > E_{\ell^+})$ denotes the number of lepton pairs
whose negatively charged member is more energetic in the $B$ meson rest frame 
than its positive partner.  Since $\CA$ is odd under charge conjugation 
whereas $\Br(\Bsll)$ is even, the information about the differential spectrum 
encoded into the former observable does not overlap with that contained 
within the latter.  The value for $\CA$ is most simply determined 
for charged $B^\pm$ mesons which do not suffer from complications associated 
with $B \dash \bar{B}$ mixing.  Counting only those lepton pairs 
whose invariant mass lies within the intervals specified in 
eqn.~\leppairinterval, we find 
\eqn\Asymmetry{\CA_{\rm NR} = {3.0 \times 10^{-7} \over 
\Br(\Bsll)_{\rm NR}} [ 1.2 - 1.0 R_7 + 4.0 R_Y - 2.6 R_Z] R_Y.}
This expression yields 7\% and 10\% for $\Bsee$ and $\Bsmm$ in the
Standard Model, respectively.

	Since deviations of $R_Y$ and $R_Z$ from unity over the allowed MSSM 
parameter space are small, supersymmetric effects in the $\Bsll$ channel 
critically depend upon the ratio $R_7$.  In order to isolate the FCNC 
mode's sensitivity to this quantity, it is instructive to first artificially 
set $R_7 =1$ everywhere throughout the region of MSSM parameter space allowed 
by all constraints.  We then find that supersymmetric effects upon $\Bsll$ are 
quite similar to those for $\kpiee$.  Nonresonant $\Bsll$ branching ratios are 
suppressed by at most 10\% relative to their Standard Model values when 
GUT-scale universality is imposed, and changes in lepton asymmetries 
are even smaller.  If the universal soft sector assumptions are relaxed, 
$B$ meson rates and asymmetries then deviate from Standard Model predictions 
by at most 30\%.

	Of course, the value for the Wilson coefficient of the magnetic 
moment operator does not coincide in most regions of MSSM parameter space 
with its Standard Model counterpart.  When GUT-scale universality is assumed
and all but $\Bsg$ constraints are imposed, we find that $R_7$ smoothly varies 
between $-2.5$ and 3 for large values of $\tan\beta$ and between 1.0 and 1.3 
for $\tan\beta \simeq 2.5$.  This $\tan\beta$ dependence primarily stems from 
the $1/\cos\beta = \sqrt{1 + \tan^2\beta}$ enhancement of the chargino 
interaction matrix $X_I^{U_R}$ in \XandZ.  If we instead search over the 
nine-dimensional parameter space discussed in section~3, we find that $R_7$ 
ranges at the factor of two level from $-\tan\beta$ to $\tan\beta$.  This 
potentially large $R_7$ variation underscores the stringent nature of the 
limits in \Rsevenlimits\ set by the CLEO $\Bsg$ observation.  

	It is useful to separately consider two different scenarios for 
$\Bsll$ decay which depend upon the sign of $R_7$:

\bigskip\noindent
$\quad {\rm Case} \sp 1.$ We accept all points in the MSSM parameter space 
for which $0.4 < R_7 < 1.2$.  The extremal MSSM values we then find for our 
$\Bsll$ observables are displayed in table~II as fractions of their 
Standard Model counterparts.  Looking at the entries in the table, we see that 
sensitivity of the nonresonant branching ratios to supersymmetry is fairly 
minimal.   The reason for the slight variation can be traced to the 
coefficients of the $R_Y$ and $R_Z$ terms in \eebr\ and \mumubr\ which are 
considerably larger than those for the $R_7$ terms.  Since $R_Y$ and $R_Z$ do 
not vary significantly from unity and the magnitude of $R_7$ is constrained by 
experiment, supersymmetric effects in the MSSM with a universal as well as 
more general soft sector are never highly pronounced in the nonresonant 
branching ratios.  

	On the other hand, the relatively larger coefficient of the $R_7$ term 
in eqn.~\Asymmetry\ induces greater sensitivity in asymmetry observables to 
supersymmetric deviations from the Standard Model.  In \AsymLEGOplot, we plot 
the ratio of the electron asymmetry observable $\CA(\Bsee)_{\rm NR}$ evaluated 
in the universal soft sector MSSM relative to its Standard Model value.  
This ratio is displayed in the figure as a function of $A_0(\MGUT)$ and 
$m_0(\MGUT)$ on a two-dimensional slice through the MSSM parameter space with 
$m_{\tildeW}(\mZ) = 150 \GeV$, $\tan\beta=20$ and ${\rm sgn}(\bigmu)=+1$ held 
fixed.  Looking at the LEGO plot, we see a substantial volume in parameter 
space exists where supersymmetric deviations from the Standard Model are 
sizable. 

$$ \vbox{\offinterlineskip
\def\tablerule{\noalign{\hrule}}
\hrule
\halign {\vrule#& \strut#& 
\quad\hfil#\hfil\quad& \vrule#& 
\quad\hfil#\hfil\quad& \vrule#& 
\quad\hfil#\hfil\quad& \vrule#& 
\quad\hfil#\hfil\quad& \vrule#& 
\quad\hfil#\hfil\quad& \vrule# \cr
height10pt && \omit && \multispan3 && \multispan3 \hfil & \cr
&& \omit && \multispan3 \hfil MSSM with universal \hfil && \multispan3 
\hfil MSSM with relaxed \hfil &\cr
&& Observable && \multispan3 \hfil soft sector \hfil && \multispan3 \hfil 
soft sector assumptions \hfil &\cr
%height2pt && \omit && \multispan3 && \multispan3 \hfil & \cr
height10pt && \omit && \multispan 7 \hrulefill & \cr
%height2pt && \omit && \omit && \omit && \omit && \omit & \cr
height10pt && \omit && minimal && maximal && minimal && maximal &\cr
\tablerule
height10pt && \omit && \omit && \omit && \omit && \omit & \cr
&& $\Br(\Bsee)_{\rm NR}$ && 77\% && 107\% && 72\% && 119\% &\cr
height10pt && \omit && \omit && \omit && \omit && \omit & \cr
&& $\CA(\Bsee)_{\rm NR} $ && 80\% && 170\% && 81\% && 196\% & \cr
height10pt && \omit && \omit && \omit && \omit && \omit & \cr
&& $\Br(\Bsmm)_{\rm NR}$ && 88\% && 102\% && 71\% && 121\% & \cr 
height10pt && \omit && \omit && \omit && \omit && \omit & \cr
&& $\CA(\Bsmm)_{\rm NR} $ && 85\% && 148\% && 86\% && 164\% & \cr
height10pt && \omit && \omit && \omit && \omit && \omit & \cr
\tablerule }} $$
\centerline{Table II.  Extremal MSSM values of $\Bsll$ observables}
\centerline{as fractions of their Standard Model counterparts.}

\bigskip\noindent
$\quad {\rm Case} \sp 2.$ We accept all points in the MSSM parameter space 
for which $-4.2 < R_7 < -2.4$.  Scans over the parameter space of the MSSM 
with GUT scale universality then reveal that $R_7$ never dips below $-2.5$.  
This limiting value coincides with the lower end of the experimentally 
permissible $\Bsg$ partial width range.  It is important to note that 
$\Gamma(\Bsee)$ is strongly correlated with $\Gamma(\Bsg)$ since a large short 
distance contribution to $\Bsee$ comes from the small dielectron mass region 
where the intermediate photon is only slightly off-shell.  Maximal suppression 
of the radiative rate therefore leads to a 10\% suppression of the 
semielectronic rate.  On the other hand, $\Gamma(\Bsmm)$ and $\Gamma(\Bsg)$ are 
anticorrelated for $R_7 < 0$ due to the $C_7^\eff \Re (C_9^\eff(\shat))$ term 
in \diffrate.  The semimuonic branching fraction increases by 20\% when 
$R_7 = -2.5$.  More importantly, sizable signals of supersymmetry can be 
detected in the asymmetry observable.  $\CA_{\rm NR}$ increases relative to its 
Standard Model value by factors of 3.6 and 2.6 for $\Bsee$ and $\Bsmm$ when
$R_7 \simeq -2.5$.

	If we again relax the universal soft sector assumptions, we find 
many points in the more general nine-dimensional parameter space where $R_7$ 
drops down to the lower end of its experimentally allowed range.
The nonresonant $\Bsee$ and $\Bsmm$ branching ratios are then enhanced by 
up to 90\% and 110\%.  Lepton asymmetries are also enhanced by approximately a 
factor of $3$ compared to Standard Model expectations when $R_7$ belongs to
the allowed negative range.
\foot{It is interesting to note that deviations of $R_7$ from unity in Case~2
can induce larger variations in $\Gamma(\Bsll)_\MSSM/\Gamma(\Bsll)_\SM$ than 
in $\Gamma(\Bsg)_\MSSM/\Gamma(\Bsg)_\SM$.}

\bigskip

	The two different scenarios we have investigated for the impact of 
supersymmetry upon $\Bsll$ decay clearly have different phenomenological 
implications.  The likelihood that this FCNC process will display interesting 
evidence for supersymmetry or else usefully constrain the MSSM parameter space 
strongly depends upon the value for $R_7$.  After more precise experimental 
measurements and next-to-leading order theoretical calculations are completed 
in the near future, the allowed range for $R_7$ should shrink by about a 
factor of 3.  It is possible that the Standard Model prediction $R_7=1$ will 
then no longer be consistent with CLEO data.  Such a result would represent an 
intriguing finding. 

	At the present time, the CLEO measurement provides no means for 
determining the sign of the $b \to s \gamma$ amplitude.  However, the current 
$3.5 \sigma$ discrepancy between the Standard Model and LEP data for 
$\Gamma(Z \to b \bar{b})$ may suggest that $R_7$ is negative in the MSSM 
framework.  Large positive MSSM corrections to $\Gamma(Z \to b \bar{b})$ would 
ameliorate the conflict between theory and experiment.  As such positive 
corrections are correlated with negative values for $R_7$, our second $\Bsll$ 
scenario may be favored.  In this case, sizable deviations in $\Bsll$ 
branching ratios and lepton asymmetries from Standard Model expectations 
should hopefully be detected in the next few years.

% ==============================================================================
\newsec{Conclusions}
% ==============================================================================

	In this article, we have studied the impact of supersymmetry upon the 
FCNC processes $\kpiee$, $\Bsee$ and $\Bsmm$.  We found that the rate for 
the kaon mode does not vary from Standard Model predictions by much more than 
10\% in the minimal supersymmetric extension with a universal soft breaking 
sector.  Qualitatively similar results hold for the $B$ meson nonresonant 
branching fraction.  Since the premise underlying the MSSM with GUT scale 
universality is not necessarily realized in nature, we have also considered 
these transitions in a more general class of models in which certain soft 
sector assumptions were relaxed.  We then uncovered regions in a 
nine-dimensional parameter space where $\Br(\Bsll)_{\rm NR}$ is significantly 
enhanced relative to its Standard Model value.  Charge conjugation odd lepton 
asymmetries can exhibit even larger deviations from Standard Model 
expectations.  So signals of supersymmetry could be detected in $\Bsll$ decay 
in the next few years.

	The general approach which we have followed in this paper to 
investigate supersymmetric contributions to a particular class of rare modes 
can be applied to several other interesting processes which might reveal 
larger discrepancies with the Standard Model.  For example, our analysis 
of semileptonic $d_i \to d_j \ell^+ \ell^-$ decay can readily be extended to 
its neutrino analogue $d_i \to d_j \nu \bar{\nu}$.  Similar methods can also 
be used to look for weak scale supersymmetry in $K \to \mu^+\mu^-$, $B \to 
\tau^+ \tau^-$ and $B \dash \bar{B}$ mixing.  Theoretical and experimental 
study of all these processes will help to constrain whatever physics lies 
beyond the Standard Model which is still waiting to be discovered.

\bigskip\bigskip
\centerline{\bf Acknowledgments}
 
        We thank P. Cooper, A. Ioannissian, T. Mannel, J. Rosiek and Y. Wah
for helpful discussions.  We are grateful to Z. Ligeti and S. Pokorski for
offering useful comments on the manuscript.  DW would also like to thank A.
Salathe for extensive assistance with programming.  Finally, we thank G.
Giudice for facilitating a comparison between our matching results and those 
reported in ref. 17.
 
\vfill\eject

% ==============================================================================
\appendix{A}{MSSM $\pmb{b \to s \gamma}$ matching conditions}
% ==============================================================================

	We list below the $W$-scale matching contributions to the 
coefficient $C_7$ of the magnetic moment operator in the $\Delta B=1$ 
effective Hamiltonian which arise from one-loop MSSM diagrams:

\noindent
Standard Model graphs: 
\eqn\Bsgmatchone{\d C_7 = {\xt \over 4} f_1(\xt)} 

\noindent
Graphs with charged Higgs loops:
\eqn\Bsgmatchtwo{\eqalign{
\d C_7 &= {1 \over 6} \left\{ \half {\mtsq \over \mhsq}
\cot^2 \beta \, f_1\bigl({\mtsq \over \mhsq} \bigr) + f_2 \bigl({\mtsq \over 
\mhsq} \bigr) \right\} \cr }}

\noindent
Graphs with chargino loops:
\eqn\Bsgmatchthree{\eqalign{
\delta C_7 &= {1 \over 3 g_2^2 \KManglesb} \sum_{\A = 1}^6 \sum_{\I = 1}^2 
{\mW^2 \over \mchI^2} \cr
& \qquad \times \Bigl\{ -\half (X_\I^\UL)^\dagger_\twoA 
 (X_\I^\UL)_\Athree f_1 \Bigl({\musqA^2 \over \mchI^2} \Bigr) 
+ (X_\I^\UL)^\dagger_\twoA (X_\I^\UR)_\Athree 
  {\mchI \over m_b} f_2 \Bigl({\musqA^2 \over \mchI^2} \Bigr) \Bigr\} \cr }}

\noindent
Graphs with neutralino loops:
\eqn\Bsgmatchfour{\eqalign{
\delta C_7 &= - {1 \over 3 g_2^2 \KManglesb} \sum_{\A = 1}^6 \sum_{\I = 1}^4 
{\mW^2 \over \mneI^2} \cr
& \qquad \times \Bigl\{ \half (Z_\I^\DL)^\dagger_\twoA 
 (Z_\I^\DL)_\Athree f_3 \Bigl({\mdsqA^2 \over \mneI^2} \Bigr) 
+ (Z_\I^\DL)^\dagger_\twoA (Z_\I^\DR)_\Athree 
  {\mneI \over m_b} f_4\Bigl({\mdsqA^2 \over \mneI^2} \Bigr) \Bigr\} \cr }}

\noindent
Graphs with gluino loops:
\eqn\Bsgmatchfive{\eqalign{
\delta C_7 &= {4 g_3^2 \over 9 g_2^2 \KManglesb} 
\sum_{\A = 1}^6 {\mW^2 \over \mgluino^2} \cr
& \qquad \times \Bigl\{ - (\Gamma^\DL)^\dagger_\twoA 
 (\Gamma^\DL)_\Athree f_3 \Bigl({\mdsqA^2 \over \mgluino^2} \Bigr) 
+ 2 (\Gamma^\DL)^\dagger_\twoA (\Gamma^\DR)_\Athree 
  {\mgluino \over m_b} f_4\Bigl({\mdsqA^2 \over \mgluino^2} \Bigr) \Bigr\} 
\cr }}

\noindent
The one-loop integral functions which enter into these matching conditions 
are given by 
\eqn\gfunctions{\eqalign{
f_1(x) &= {-7 + 5 x + 8x^2  \over 6(1-x)^3} - {2 x - 3 x^2 \over (1-x)^4} 
  \log x \cr
f_2(x) &= {3x-5x^2 \over 2(1-x)^2} + {2x-3x^2 \over (1-x)^3} 
  \log x \cr
f_3(x) &= {2+ 5 x - x^2 \over 6(1-x)^3} + {x \over (1-x)^4} \log x \cr
f_4(x) &= {1 + x \over 2 (1-x)^2} + {x \over (1-x)^3} \log x. \cr}}

\vfill\eject

% ==============================================================================
\appendix{B}{MSSM $\pmb{\sbar \to \dbar e^+ e^-}$ and $\pmb{b \to s e^+ e^-}$ 
matching conditions}
% ==============================================================================

	We tabulate below the $W$-scale matching contributions to the 
$Y^\SUSY$ and $Z^\SUSY$ parameters which appear in the Wilson coefficients 
$y_{7V}$ and $y_{7A}$ of the strange sector semileptonic operators $Q_{7V}$ 
and $Q_{7A}$ in eqn.~\coefficients.  These same formulae hold for Wilson 
coefficients of analogous operators in the bottom sector.  The KM matrix label 
$q$ and numerical index $i$ respectively equal $q=d$, $i=1$ and $q=b$, $i=3$ 
for $\sbar \to \dbar e^+ e^-$ and $b \to s e^+ e^-$ decay.

\noindent
Z-penguin and bubble graphs with charged Higgs loops:
\eqn\kpieematchone{
\delta Y^\SUSY = \delta Z^\SUSY = -{1 \over 8} 
\cot^2 \beta \, \xt f_5\bigl({\mtsq \over \mhsq} \bigr)} 

\noindent
$\gamma$-penguin and bubble graphs with charged Higgs loops:
\eqn\kpieematchtwo{\eqalign{
\delta Y^\SUSY &= 0 \cr
\delta Z^\SUSY &= -{1 \over 72} \cot^2 \beta \, f_6\bigl({\mtsq \over \mhsq} 
\bigr) \cr }}

\noindent
Z-penguin and bubble graphs with chargino loops:
\eqn\kpieematchthree{\eqalign{
\delta Y^\SUSY &= \delta Z^\SUSY = {1 \over 2 g_2^2 \KMangles}
\sum_{\A,\B = 1}^6 \sum_{\I,\J = 1}^2 (X_\I^\UL)^\dagger_\twoA (X_\J^\UL)_\Bi
\cr 
& \qquad \times \Bigl\{ c_2(\mchI^2, \musqA^2, \musqB^2)
(\Gamma^\UL {\Gamma^\UL}^\dagger)_{\A\B} \, \dIJ
- c_2(\musqA^2, \mchI^2, \mchJ^2)  \dAB V^*_{\I1} V_{\J1} \cr
& \qquad + \half \mchI \mchJ c_0(\musqA^2, \mchI^2, \mchJ^2) \dAB
U_{\I1} U^*_{\J1} \Bigr\} \cr }}

\noindent
$\gamma$-penguin and bubble graphs with chargino loops:
\eqn\kpieematchfour{\eqalign{
\delta Y^\SUSY &= 0 \cr
\delta Z^\SUSY &= {1 \over 36 g_2^2 \KMangles} \sum_{\A=1}^6 \sum_{\I=1}^2
{\mWsq \over \musqA^2} (X_\I^\UL)^\dagger_\twoA (X_\I^\UL)_\Ai 
f_7\bigl( {\mchI^2 \over \musqA^2} \bigr) \cr }}

\noindent
Z-penguin and bubble graphs with neutralino loops:
\eqn\kpieematchfive{\eqalign{
& \delta Y^\SUSY = \delta Z^\SUSY = {1 \over 2 g_2^2 \KMangles} 
\sum_{\A,\B=1}^6 \sum_{\I,\J=1}^4
(Z_\I^\DL)^\dagger_\twoA (Z_\J^\DL)_\Bi \cr 
& \quad \times \Bigl\{ c_2(\mneI^2, \mdsqA^2, \mdsqB^2) 
(\Gamma^\DR {\Gamma^\DR}^\dagger)_{\A\B} \, \dIJ 
- c_2(\mdsqA^2, \mneI^2, \mneJ^2) 
\dAB (N^*_{\I3} N_{\J3} - N^*_{\I4} N_{\J4} ) \cr
& \quad - \half \mneI \mneJ c_0(\mdsqA^2, \mneI^2, \mneJ^2) 
\dAB (N_{\I3} N^*_{\J3} - N_{\I4} N^*_{\J4} ) \Bigl\} \cr }}

\noindent
$\gamma$-penguin and bubble graphs with neutralino loops:
\eqn\kpieematchsix{\eqalign{
\delta Y^\SUSY &= 0 \cr
\delta Z^\SUSY &= - {1 \over 216 g_2^2 \KMangles} \sum_{\A=1}^6 \sum_{\I=1}^4
{\mWsq \over \mdsqA^2} (Z_\I^\DL)^\dagger_\twoA (Z_\I^\DL)_\Ai 
f_8\bigl( {\mneI^2 \over \mdsqA^2} \bigr) }} 

\noindent
Z-penguin and bubble graphs with gluino loops:
\eqn\kpieematchseven{\eqalign{
\delta Y^\SUSY = \delta Z^\SUSY &= {4 g_3^2 \over 3 g_2^2 \KMangles} 
\sum_{\A,\B=1}^6
(\Gamma^\DL)^\dagger_\twoA (\Gamma^\DL)_\Bi c_2(\mgluino^2, \mdsqA^2, 
\mdsqB^2) (\Gamma^\DR {\Gamma^\DR}^\dagger)_{\A\B} \cr }}

\noindent
$\gamma$-penguin and bubble graphs with gluino loops:
\eqn\kpieematcheight{\eqalign{
\delta Y^\SUSY &= 0 \cr
\delta Z^\SUSY &= - {g_3^2 \over 81 g_2^2 \KMangles} \sum_{\A=1}^6
{\mWsq \over \mdsqA^2} 
(\Gamma^\DL)^\dagger_\twoA (\Gamma^\DL)_\Ai f_8\bigl( {\mgluino^2 \over 
\mdsqA^2} \bigr) \cr }}

\noindent
Chargino box graph: 
\foot{Our chargino and neutralino box graph matching condition results differ 
from those reported in ref.~\Bertolini\ by overall signs.}
\eqn\kpieematchnine{\eqalign{
\delta Y^\SUSY &= {\mWsq \over g_2^2 \KMangles} 
\sum_{\A=1}^6 \sum_{\I,\J=1}^2
(X_\I^\UL)^\dagger_\twoA (X_\J^\UL)_\Ai d_2(\mchI^2, \mchJ^2, \musqA^2, 
\msneutrinoone^2) V^*_{\I1} V_{\J1} \cr
\delta Z^\SUSY &= 0 \cr }} 

\noindent
Neutralino box graphs: 
\eqn\kpieematchten{\eqalign{
& \delta Y^\SUSY = 2 \sinsq\;\delta Z^\SUSY + {\mWsq \over 2 g_2^2 \KMangles} 
\sum_{\A=1}^6 \sum_{\I,\J=1}^4
(Z_\I^\DL)^\dagger_\twoA (Z_\J^\DL)_\Ai \cr
& \quad \times \Bigl\{ d_2(\mneI^2, \mneJ^2, \mdsqA^2, \meslone^2) 
  (N^*_{\I2} + \tan\theta N^*_{\I1})(N_{\J2} + \tan\theta N_{\J1}) \cr 
& \qquad + \half \mneI \mneJ  d_0(\mneI^2, \mneJ^2, \mdsqA^2, \meslone^2) 
  (N_{\I2} + \tan\theta N_{\I1})(N^*_{\J2} + \tan\theta N^*_{\J1}) \Bigr\} \cr
& \delta Z^\SUSY = {\mWsq \over g_2^2 \KMangles} 
\sum_{\A=1}^6 \sum_{\I,\J=1}^4
(Z_\I^\DL)^\dagger_\twoA (Z_\J^\DL)_\Ai \sec^2 \theta \cr
& \quad \times \bigl[ d_2(\mneI^2, \mneJ^2, \mdsqA^2, \meslfour^2) 
  N^*_{\I1} N_{\J1} 
  + \half\mneI\mneJ d_0(\mneI^2,\mneJ^2,\mdsqA^2,\meslfour^2) N_{\I1} N^*_{\J1}
  \bigr] \cr }}
 
\vfill\eject

The one-loop integral functions which appear within these MSSM matching 
conditions are given by 
\eqn\functions{\eqalign{
f_5(x) &= {x \over 1-x} + {x \over (1-x)^2} \log x \cr
f_6(x) &= {38x-79x^2+47 x^3 \over 6(1-x)^3} + {4x-6x^2+3x^4 \over (1-x)^4} 
\log x \cr
f_7(x) &= {52-101x+43x^2 \over 6(1-x)^3} + {6-9x+2x^3 \over (1-x)^4} 
\log x \cr
f_8(x) &= {2-7x+11x^2 \over (1-x)^3} + {6x^3 \over (1-x)^4} \log x \cr
%
%\eqn\Cintegrals{\eqalign{
%
c_0(m_1^2, m_2^2, m_3^2) &= -\Bigl[
  {m_1^2 \, \displaystyle{\log{m_1^2 \over \mu^2}} \over
  (m_1^2-m_2^2)(m_1^2-m_3^2)}
+ (m_1 \leftrightarrow m_2)
+ (m_1 \leftrightarrow m_3) \Bigr] \cr 
c_2(m_1^2, m_2^2, m_3^2) &= {3 \over 8} - {1 \over 4} \Bigl[
  {m_1^2 \, \displaystyle{\log{m_1^4 \over \mu^2}} \over
  (m_1^2-m_2^2)(m_1^2-m_3^2)}
+ (m_1 \leftrightarrow m_2)
+ (m_1 \leftrightarrow m_3) \Bigr] \cr
%
%\eqn\Dintegrals{\eqalign{
%
d_0(m_1^2, m_2^2, m_3^2, m_4^2) &= \cr
- \Bigl[ &
{m_1^2 \, \displaystyle{\log{m_1^2 \over \u^2}} \over
  (m_1^2-m_2^2)(m_1^2-m_3^2)(m_1^2-m_4^2)} 
+ (m_1 \leftrightarrow m_2)
+ (m_1 \leftrightarrow m_3) 
+ (m_1 \leftrightarrow m_4) \Bigr] \cr 
& \cr
d_2(m_1^2, m_2^2, m_3^2, m_4^2) &= \cr
- {1 \over 4} \Bigl[ & 
{m_1^4 \, \displaystyle{\log{m_1^2 \over \u^2}} \over
  (m_1^2-m_2^2)(m_1^2-m_3^2)(m_1^2-m_4^2)} 
+ (m_1 \leftrightarrow m_2)
+ (m_1 \leftrightarrow m_3) 
+ (m_1 \leftrightarrow m_4) \Bigr]. \cr}}
\noindent
All dependence upon the renormalization scale $\mu$ cancels out from 
the total supersymmetric matching conditions in $Y^\SUSY$ and $Z^\SUSY$.

\listrefs
\listfigs
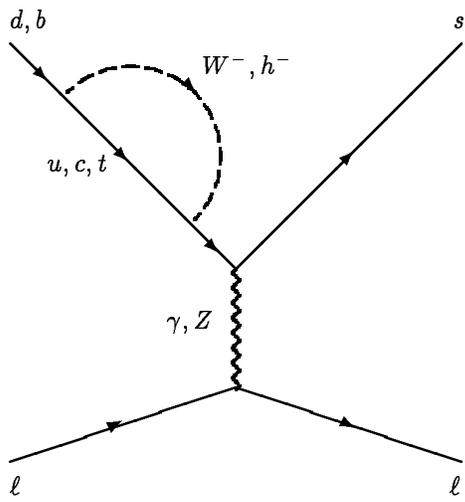
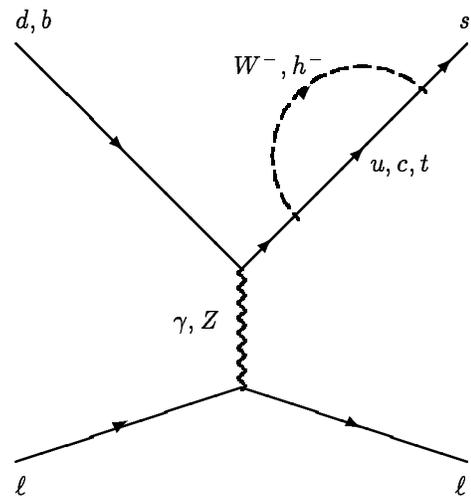
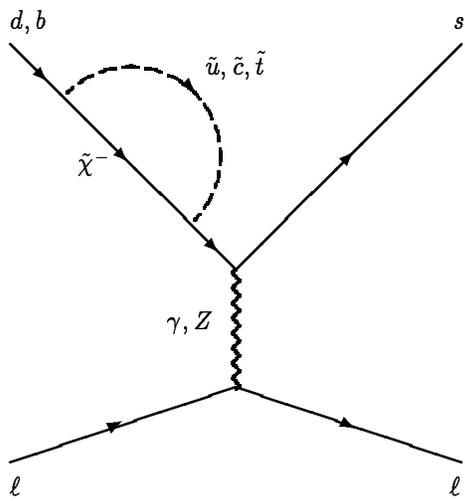
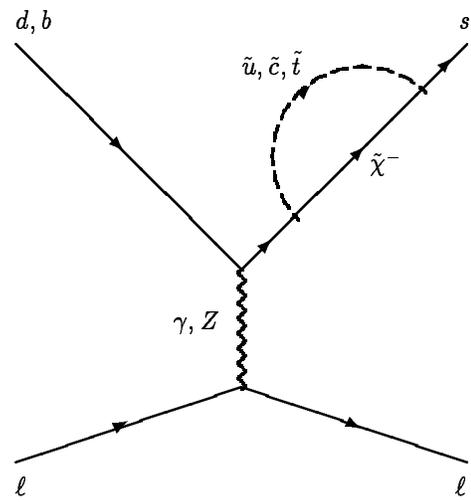
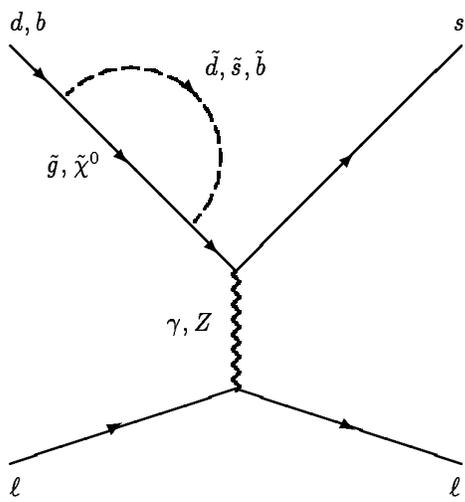
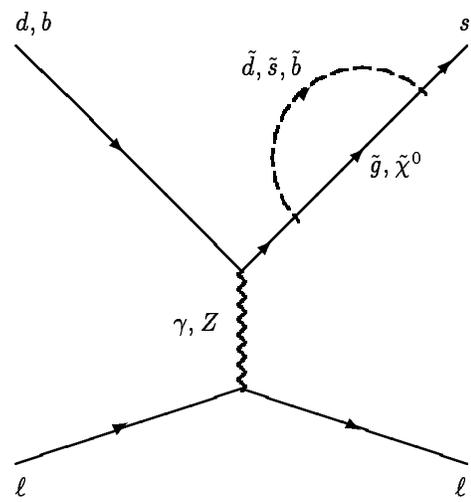
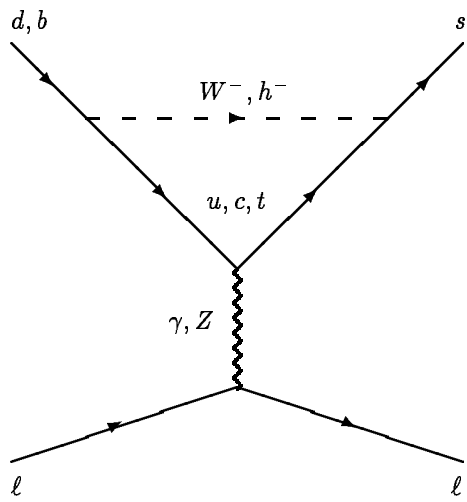
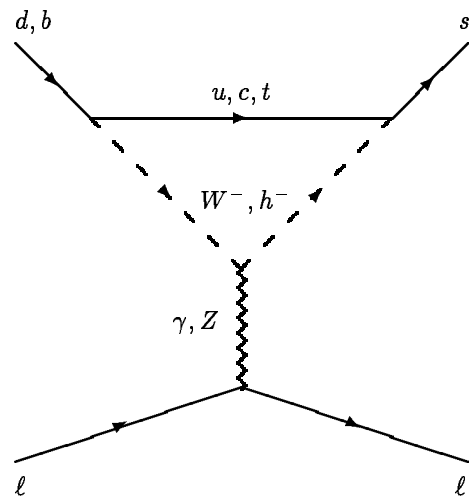
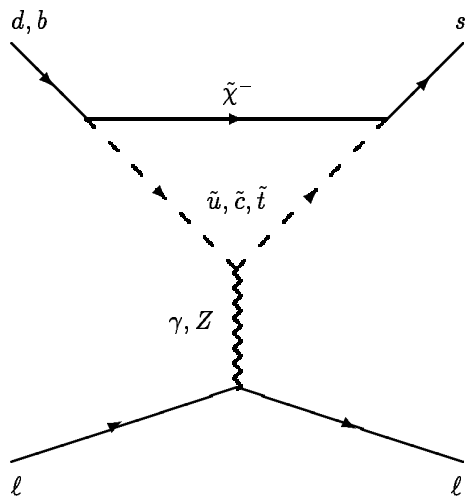
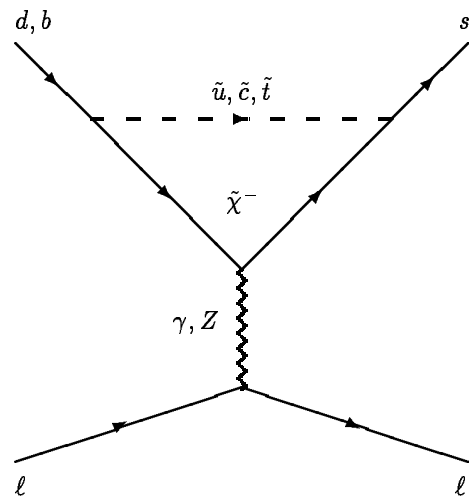
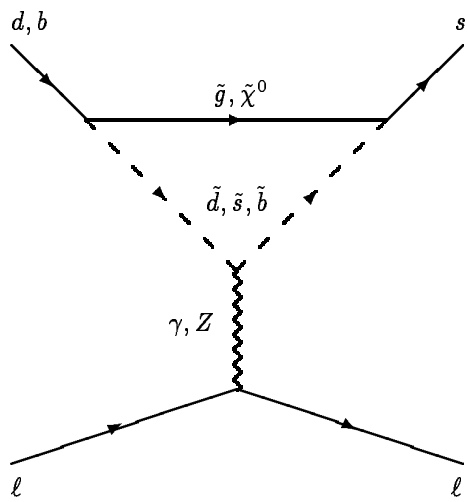
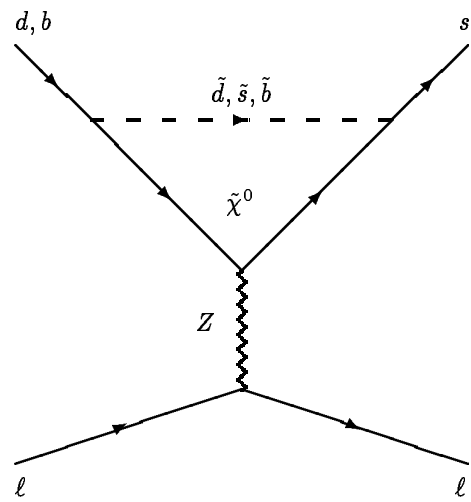
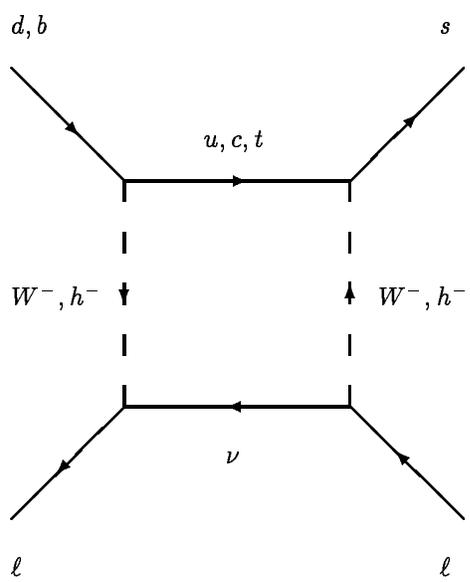
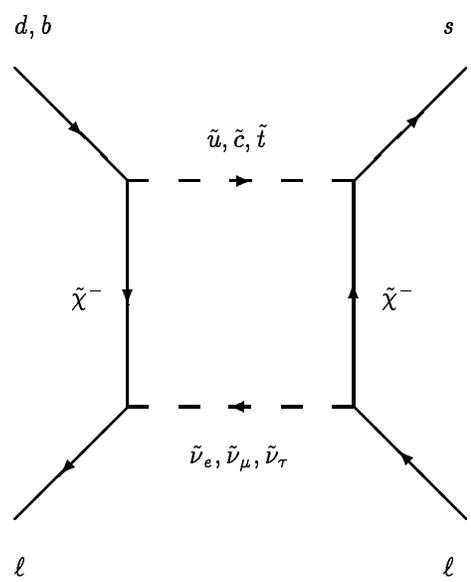
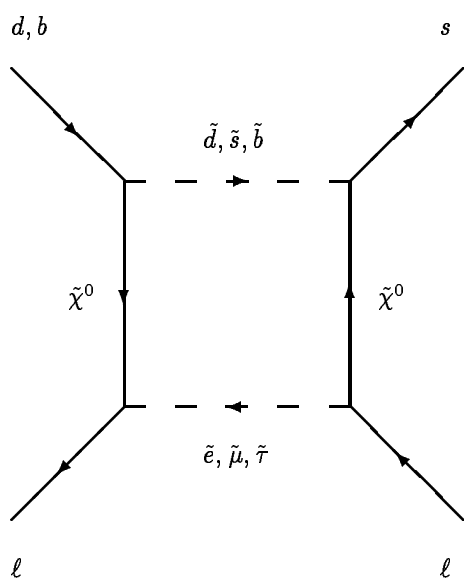
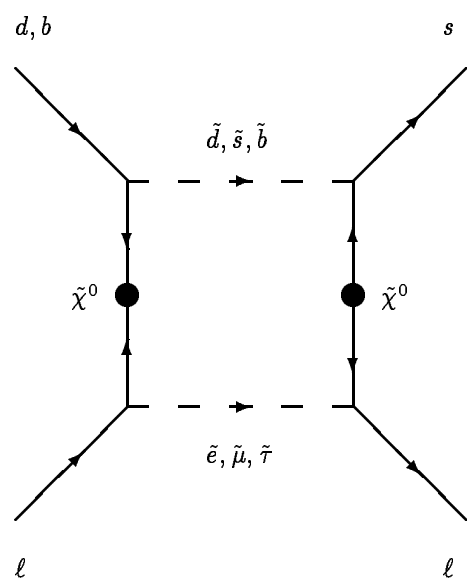
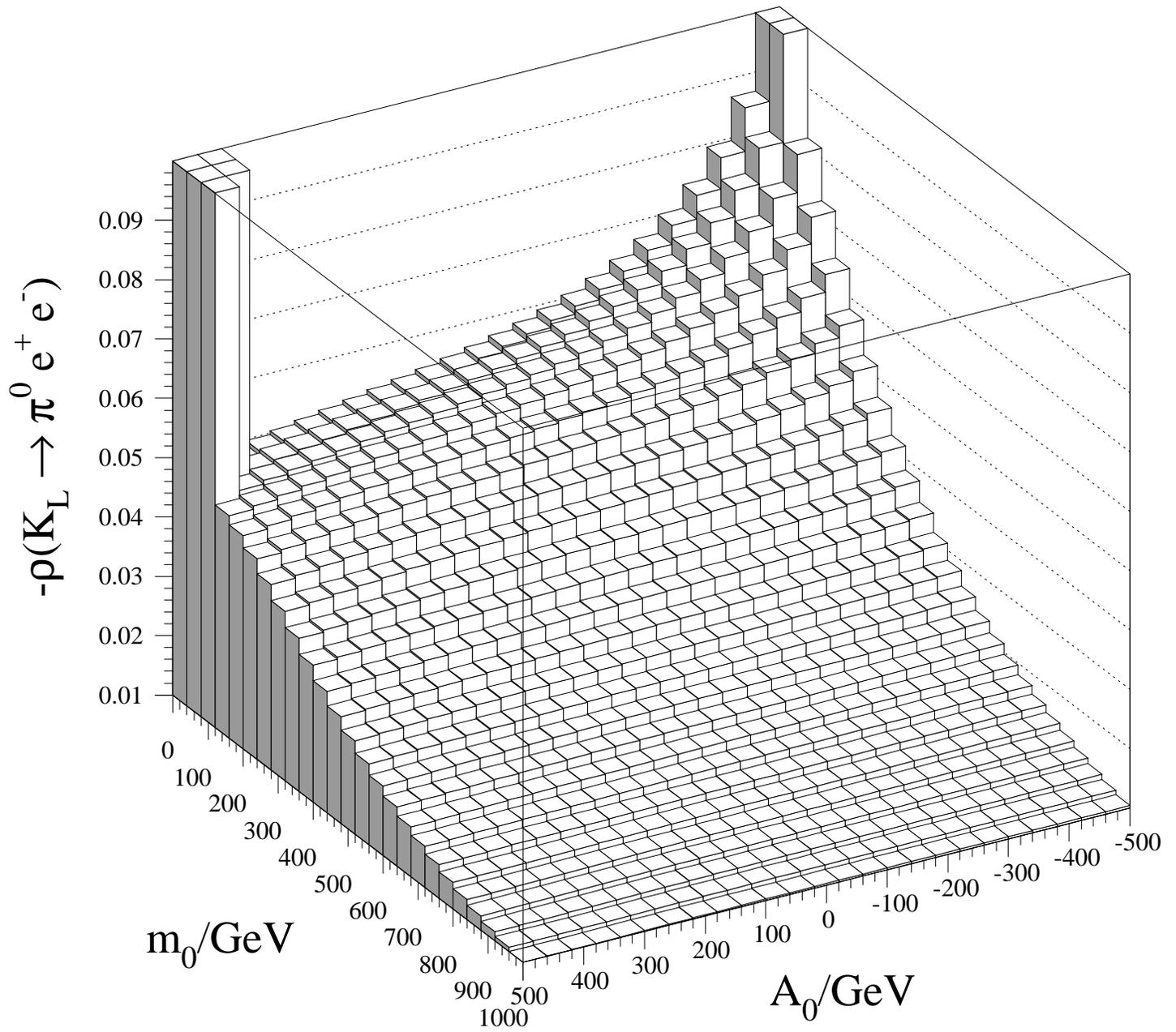
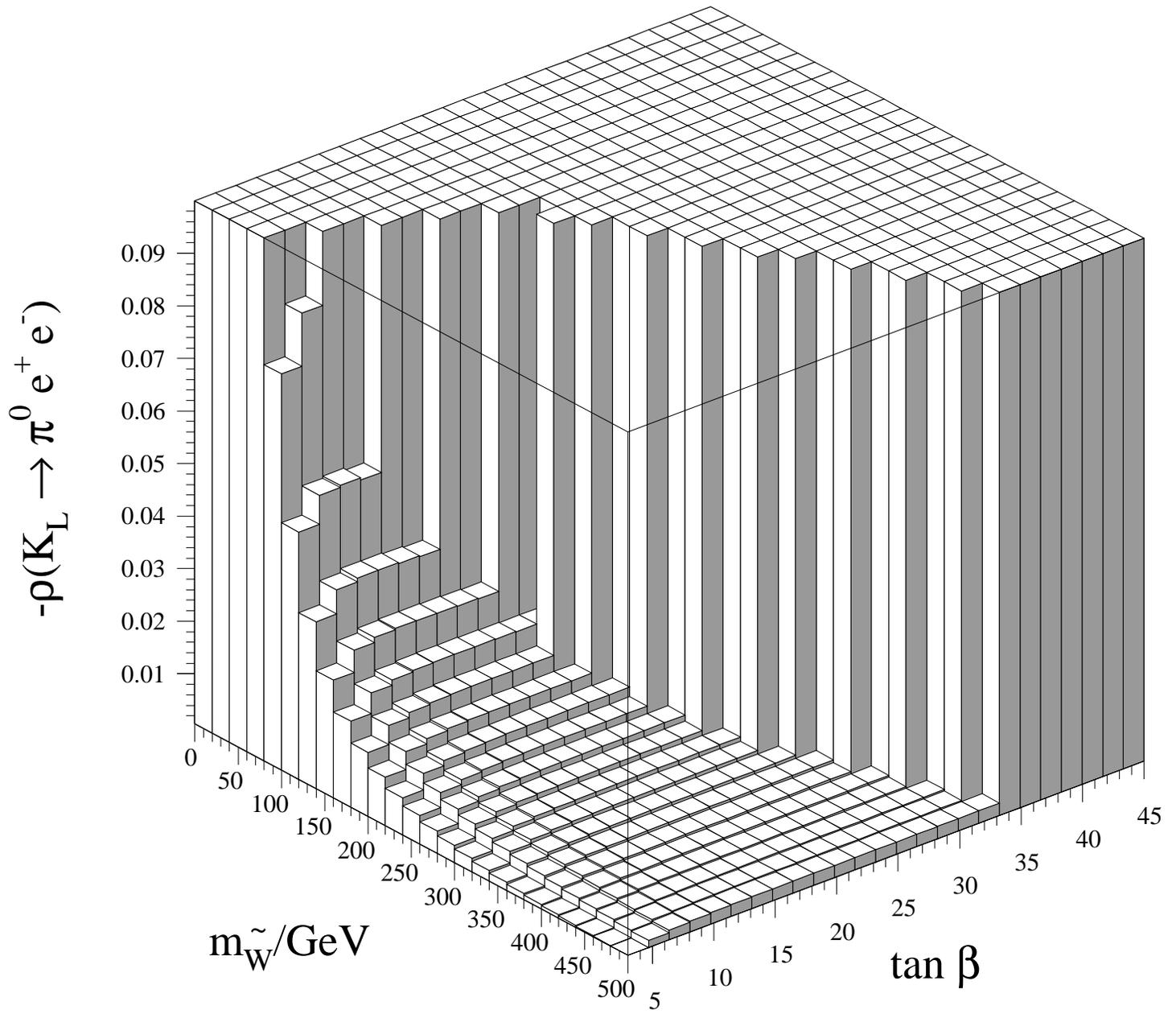
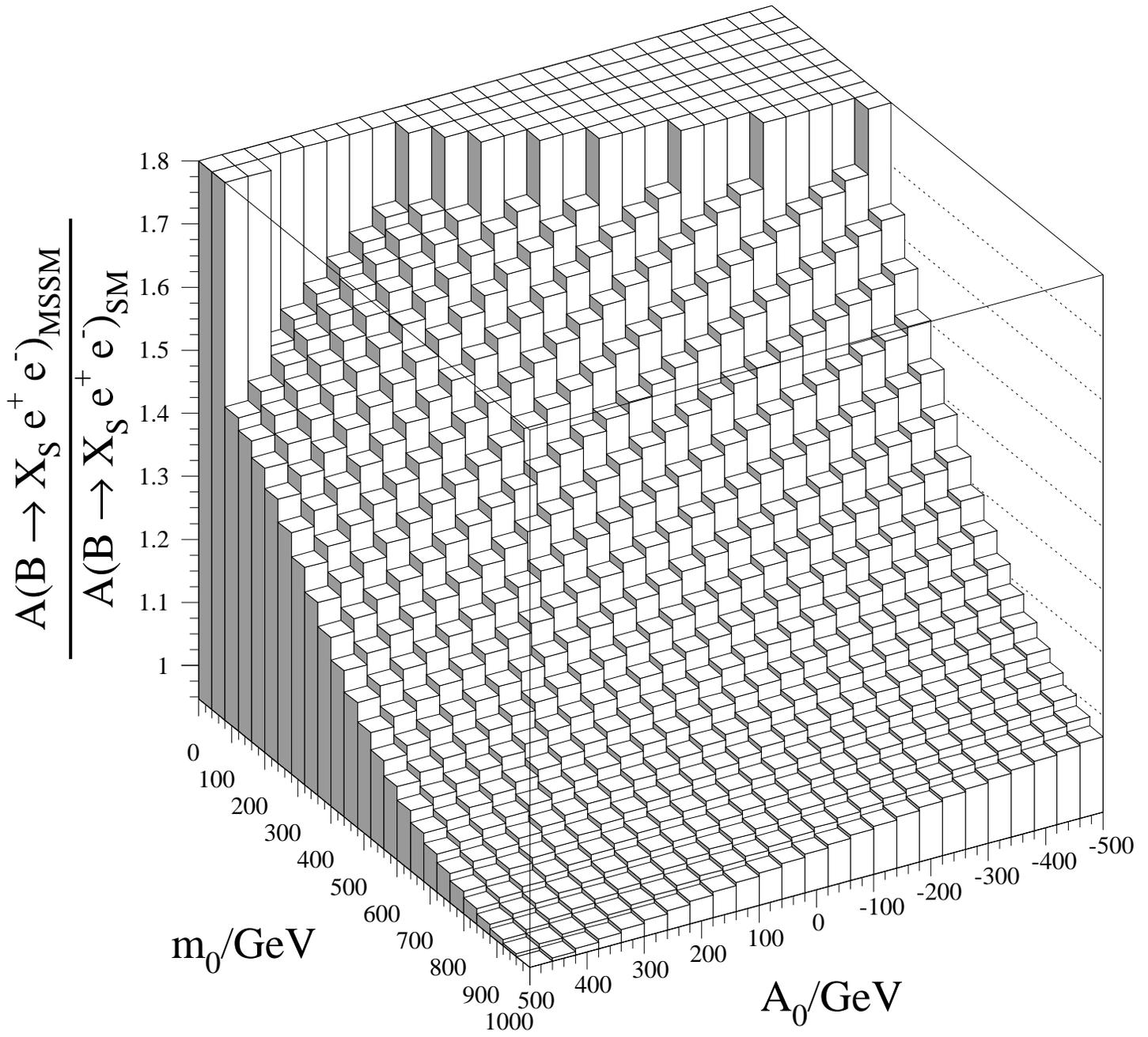
\bye